\documentclass[24pt]{article}
\usepackage{geometry}
\usepackage{xcolor}
\usepackage{a4}
\usepackage{alltt}
\usepackage{graphicx}
\usepackage{epsf}
\usepackage{amsmath}
\usepackage{amssymb}
\usepackage{cite}
\usepackage{hyperref}
\usepackage{blindtext}
\usepackage[toc,page]{appendix}

\newcommand{\be}{\begin{equation}}
\newcommand{\ee}{\end{equation}}

\newcommand{\Rmnum}[1]{\expandafter\@slowromancap\romannumeral #1@}
\newcommand{\bea}{\begin{eqnarray}}
\newcommand{\eea}{\end{eqnarray}}
\begin{document}
\title{\bf Cosmological bounce and some other solutions in exponential gravity} 
\author{Pritha Bari, Kaushik Bhattacharya, Saikat Chakraborty 
\thanks{ E-mail:~ pribari@iitk.ac.in, kaushikb@iitk.ac.in, snilch@iitk.ac.in} \\ 
\normalsize Department of Physics, Indian Institute of Technology,\\ 
\normalsize Kanpur 208016, India}
\maketitle
\begin{abstract}
  In this work we present some cosmologically relevant solutions using
  the spatially flat Friedmann-Lemaitre-Robertson-Walker (FLRW)
  spacetime in metric $f(R)$ gravity where the form of the
  gravitational Lagrangian is given by $\frac{1}{\alpha}e^{\alpha
    R}$. In the low curvature limit this theory reduces to ordinary
  Einstein-Hilbert Lagrangian together with a cosmological constant
  term. Precisely because of this cosmological constant term this
  theory of gravity is able to support nonsingular bouncing solutions
  in both matter and vacuum background. Since for this theory of
  gravity $f^{\prime}$ and $f^{\prime\prime}$ is always positive, this
  is free of both ghost instability and tachyonic
  instability. Moreover, because of the existence of the cosmological
  constant term, this gravity theory also admits a de-Sitter
  solution. Lastly we hint towards the possibility of a new type of
  cosmological solution that is possible only in higher derivative
  theories of gravity like this one.
\end{abstract}
\section{Introduction}

Investigation of non-singular bouncing cosmological solutions to
Einstein field equations has a history that dates back to first half
of the twentieth century, and can be attributed to various works of
Lemaitre, Tolman, Friedmann and even Einstein himself(see,
e.g. ref.\cite{Kragh:2013dva} for early histories of bouncing
cosmology). But it was then largely considered just as an alternative
solution to Einstein equations and did not have any physical
motivation as such. The most accepted paradigm about the universe
evolution then was the big-bang paradigm, which is based upon the
existence of a curvature singularity in the past. However, big-bang
singularity was plagued with several other problems. Inflationary
scenario came into picture in the later half of the nineteenth century
as a very promising candidate to solve these problems. This paradigm
was pioneered by the works of Guth \cite{Guth:1980zm} and
Linde \cite{Linde:1981mu}. There are many models in the literature so
far that realize an inflationary scenario (see \cite{Martin:2013tda}
for a comprehensive review of all the inflationary models).

Although inflationary cosmology is highly successful in explaining various
features of the early universe, it is still plagued with the issue of
singularity\footnote{The reader can consult Ref.~\cite{Borde:1996pt}
  for a review on singularities in inflationary cosmology.}, which, by
definition, is a state of physical lawlessness. When we try to
describe our universe with the available physical theories, we usually
do not want a singularity to come into the picture. This was the
physical motivation which refuelled the interest in nonsingular
bouncing scenarios. There are various ways to realize a bouncing
scenario(see \cite{Novello:2008ra} or \cite{Battefeld:2014uga} for a
comprehensive review).

If one tries to realize a bouncing solution for spatially flat FLRW
metric in general relativity (GR), one needs to invoke null energy
condition (NEC) violating matter components like ghost fields, ghost
condensates or Galileons. If one does not wish to invoke such exotic
matter components and still wants to realize a bouncing solution in
spatially flat FLRW metric, then he/she has to resort to modified
gravity. Modifications to general relativity at high curvature regime
near a curvature singularity can indeed be expected. When quantum
corrections or string theory motivated effects are taken into account,
then the effective low energy gravitational action indeed admits
higher order curvature invariant terms \cite{Buchbinder:1992rb,
  Vilkovisky:1992pb}. The simplest of such modifications is when the
correction terms depend only on the Ricci scalar $R$. In such cases
the Einstein-Hilbert Lagrangian $R$ is modified to $f(R)$, a function
of $R$ (see \cite{Sotiriou:2008rp}, \cite{DeFelice:2010aj},
\cite{Nojiri:2017ncd} and \cite{Nojiri:2010wj} for beautiful reviews
on $f(R)$ gravity).

There has been many attempts to realize bouncing scenario in $f(R)$
theories of gravity. It was first pointed out in Ref.\cite{ruz2} that
$R+\alpha R^2$ gravity with a negative $\alpha$ can give rise to a
bouncing scenario in spatially flat FLRW metric. Some authors have
even used $f(R)$ gravity to tackle the issues related to cosmological
bounce and a cyclic universe \cite{Cai:2011bs,
  Saridakis:2018fth}. Bouncing cosmology for quadratic and cubic
polynomial $f(R)$ gravity theory was more recently worked out in
\cite{Paul:2014cxa} and \cite{Bhattacharya:2015nda} respectively, from
both Jordan frame and Einstein frame point of view. It was shown in
\cite{Paul:2014cxa} that for spatially flat FLRW metric a bounce in
the Jordan frame is never accompanied by a bounce in the Einstein
frame when hydrodynamic matter in the Jordan frame satisfies the
condition $\rho + P \ge 0$. Here $\rho$ an $P$ specifies the energy
density and pressure of matter in the Jordan frame. Working solely in
the Jordan frame, it was shown in \cite{Bamba:2013fha} that quadratic
gravity theories of the form $\lambda+R+\alpha R^2$ with a negative
$\alpha$ and monomial gravity theories of the form $R^{1+\delta}$ can
also produce bouncing cosmologies. Carloni \textit{et. al.} in
\cite{Carloni:2005ii} presented the bouncing conditions in $f(R)$
gravity and also analyzed the conditions for $R^{1+\delta},\,R+\alpha
R^m$ and $\exp{\alpha R}$ type of gravity.

Two necessary conditions for the physical viability of any $f(R)$
theory are $f'(R)>0$ and $f''(R)>0$. As we will see in a later
section, all the other $f(R)$ theories except $e^{\alpha
  R}$($\alpha>0$) that has commonly been considered in the literature
so far to realize a bounce in spatially flat FLRW metric can not have
$f'(R)>0$ and $f''(R)>0$ simultaneously for all $R$. So exponential
gravity may be the only physically viable candidate for achieving a
bounce. Carloni \textit{et. al.} in \cite{Carloni:2005ii} concludes
that this gravity theory can give rise to a bouncing solution only in
a closed FLRW universe, but as we will show later on, this theory can
also produce a bounce in the flat FLRW universe. In previous studies
exponential gravity has been used extensively used to study
cosmological inflation and late time acceleration of the universe
\cite{Elizalde:2010ts, Oikonomou:2018npe, Oikonomou:2013rba}. The
motivation of the present paper is to present a metric $f(R)$ theory
of gravity, which is free of the above mentioned instabilities, and
which can produce a successful cosmological bounce in the early
universe. The model of bounce which we present here is to be taken as
an effective theory in high energy scales as exponential $f(R)$ only
describes the system very near the bounce point. Essentially we
present a bounce mechanism and not a full description of cosmology
which includes how physics much prior to bounce is related to the
bouncing period, although our model can have a transition to low
energy GR for small values of the Ricci scalar but we presume such a
cross-over to low energy theory may require new physics.  As specified
in Ref.\cite{Cai:2016hea} the prebounce cosmology can be connected to
a particular bounce mechanism and out of various bounce mechanism
$f(R)$ bounce is one. The inclusion of higher order terms in the Ricci
scalar $R$ in the action is often motivated by two main observations:
first, adding new terms in the space-time curvature could explain the
observations typically associated to dark matter and/or dark energy,
and second, since the Einstein-Hilbert action is not renormalizable,
any consistent theory of quantum gravity is expected to contain higher
order curvature terms in the action that become important near the
Planck scale.  While $f(R)$ theory is not the most general kind of
such theory, it is one of the simplest modifications possible on GR
and are often viewed as a good first step in understanding the effect
of adding additional terms to the Einstein-Hilbert
action. Consequently we assume that the prebounce phase of our theory
will be some form of cosmological theory as discussed in
\cite{Cai:2016hea} or in Ref.\cite{Cai:2014bea}. In \cite{Cai:2014bea}
the author takes up a case where there is a bounce followed up by
inflation, in the present we do not expect inflation after bounce in
exponential gravity but exponential gravity do have an exact de Sitter
solution and we briefly discuss cosmology of the de Sitter solution
later in this article.

In this article we briefly study the nature of scalar metric
perturbations and opine briefly on the tensor perturbations through
bounce. The perturbation theory is completely presented in the Jordan
frame. It is shown that scalar perturbations do remain non-singular
near the bouncing point but there can be new instabilities in the new
system which can make some some of the modes to become
non-perturbative near the bounce. We also show that we do not expect a
higher tensor-to-scalar ratio in our bounce model.

In this paper we also present a new solution of $f(R)$ gravity theories
which admits a cosmological bounce. The new solution is related to more degrees
of freedom of $f(R)$ theories.  Unlike GR, $f(R)$ theories depend
on the second time derivative of the Hubble parameter and one can tune
the cosmological development of a model by specifying various values
of $\ddot{H}$ at some specific time. These new solutions can produce
interesting new model universes. We present a new solution in bouncing
$f(R)$ theories where the universe transits from an decelerated
expansion phase to a normal expansion phase vis a contraction
phase. The new result presented in the present paper is general and we
have given some specific examples using exponential gravity as an
example.

The material in this paper is presented in the following way. In the
next section we present the formalism of $f(R)$ theory in the two
conformal frames, the Jordan frame and the Einstein frame. The
relation between the frames is presented in the this sections. The
bouncing scenario is described in section \ref{dbscen}. Section
\ref{begs} specifies bounce in exponential gravity. In this section we
present the numerical results depicting various kinds of bounces.
Discussion on scalar metric perturbation through bounce is presented
in section \ref{ptb}. Two exact solutions of exponential gravity is
presented in section \ref{tes}. We discuss the new solutions in $f(R)$
gravity theories in section \ref{new}. The next section concludes the
article by summarizing the results obtained.
\section{The cosmological set up in the two frame}

In this section we present the formal structure of the theory we will
pursue in this article. The field equations in the two conformal frames and the formulae connecting the frames are briefly specified in this section.
\subsection{The Jordan frame}

In the Jordan frame the field equations for $f(R)$ gravity is given in
the tensorial form as follows,
\begin{eqnarray}
G_{\mu\nu}\equiv R_{\mu\nu}-\frac{1}{2}g_{\mu\nu}R &=&\frac{\kappa T_{\mu\nu}}
{f^{\prime}(R)}+g_{\mu\nu}\frac{\left[f(R)-Rf^{\prime}(R)\right]}{2f^{\prime}(R)}
\nonumber\\
&+&\frac{\nabla_{\mu}\nabla_{\nu}f^{\prime}(R)-g_{\mu\nu}\square f^{\prime}(R)}
{f^{\prime}(R)}\,.
\label{ein1}
\end{eqnarray}
For the flat FLRW metric given by 
\begin{equation}
ds^2 = -dt^2 + a^2(t)d\bar{x}^2\,,
\label{frw}
\end{equation}
and for a perfect fluid given by the energy-momentum tensor
\begin{equation}
T_{\mu \nu} = (\rho + P)u_\mu u_\nu + P g_{\mu\nu}\,,
\label{tmunu}
\end{equation}
the modified Friedmann equations for $f(R)$ gravity are:
\begin{eqnarray}
3H^2&=&\frac{\kappa}{f^{\prime}(R)} (\rho
  + \rho_{\rm eff})\,,
\label{fried}\\
3H^{2}+2\dot{H} &=&\frac{-\kappa}{f^{\prime}(R)}
(P+ P_{\rm eff})\,,
\label{2ndeqn}
\end{eqnarray}
where $\rho_{\rm eff}$ and $P_{\rm eff}$ are related to energy density
and pressure arising from curvature. They are defined by the following expressions,
\begin{eqnarray}
\rho _{\rm eff} &\equiv& \frac{Rf^{\prime}-f}{2\kappa}-\frac{3H\dot{R}
  f^{\prime \prime}(R)}{\kappa}\,,
\label{reff}\\
P_{\rm eff} &\equiv& \frac{\dot{R}^{2}f^{\prime \prime \prime} + 2H\dot{R}f^{\prime
    \prime}+ \ddot{R}f^{\prime \prime} }{\kappa} - \frac{Rf^{\prime}-f}{2\kappa}\,.
\label{peff}
\end{eqnarray}
For a barotropic fluid,
\begin{eqnarray}
P=\omega \rho\,,
\label{eqns}
\end{eqnarray}
where $\omega=0$ corresponds to dust and $\omega=\frac{1}{3}$
corresponds to radiation. It must be noted that the 4-velocity $u_\mu$
in Eq.~(\ref{tmunu}) is the normalized 4-velocity of a fluid
element. The other relevant dynamical equation is the continuity
equation for the hydrodynamic matter component
\begin{eqnarray}
\dot{\rho}+3H\rho(1+\omega)=0\,.
\label{cont}
\end{eqnarray}
\subsection{The Einstein frame}

The equivalent Einstein frame description of $f(R)$ gravity is defined
in terms of a conformally related metric
\begin{eqnarray}
\tilde{g}_{\mu \nu} = F(R)g_{\mu \nu}\,,
\label{conf}
\end{eqnarray}
where $$F(R)\equiv\frac{d f(R)}{dR}\,.$$ Let us define in
Einstein frame a scalar field $\phi$ and the potential $V(\phi)$ as
\begin{eqnarray}
  \phi \equiv \sqrt{\frac{3}{2\kappa}}\ln F\,,\,\,\,\,\,
  V(\phi)=\frac{R F-f}{2\kappa F^2}\,.
\label{vexp}
\end{eqnarray}
Gravitational dynamics in Einstein frame is affected by the dynamics
of the scalar field which comes into existence in the Einstein frame
due to the conformal transformation.  Under the conformal
transformation in Eq.~(\ref{conf}) the energy-momentum tensor
transforms as
\begin{eqnarray}
\tilde{T}^\mu_\nu = \frac{T^\mu_\nu}{F^2}\,. 
\label{conft}
\end{eqnarray}
Specifically, we have the relations
\begin{eqnarray}
\tilde{\rho}=\frac{\rho}{F^2}\,,\,\,\tilde{P}=\frac{P}{F^2}\,,\,\,
\tilde{u}_\mu \equiv \sqrt{F} u_\mu\,.
\label{nrhop} 
\end{eqnarray}
In the Jordan frame there is only one hydrodynamic matter component,
where as in the Einstein frame there is also a scalar field defined
above, which has a potential given by Eq.~(\ref{vexp}), which couples
non-minimally with the hydrodynamic matter component. Since in the
Einstein frame the gravitational theory is GR, we can write the field
equations in the tensorial form as
\begin{eqnarray}
\tilde{R}^\mu_\nu - \frac12 \delta^\mu_\nu \tilde{R}
= \kappa \left(\tilde{T}^\mu_\nu + \tilde{\mathcal T}^\mu_\nu\right)\,.
\label{neqn}
\end{eqnarray}
where $\tilde{T}_{\mu\nu}$ is the energy-momentum tensor of the
hydrodynamic fluid in the Einstein frame
\begin{eqnarray}
\tilde{T}_{\mu\nu} = (\tilde{\rho} + \tilde{P})\tilde{u}_\mu 
\tilde{u}_\nu + \tilde{P} \tilde{g}_{\mu\nu}\,,
\label{tmunup}
\end{eqnarray}
and $\tilde{\mathcal T}^\mu_\nu$ is the energy-momentum tensor of the
scalar field $\phi$. Here $\tilde{R}^\mu_\nu$ is the corresponding
Ricci tensor in the Einstein frame. We can recast the conformally
related Einstein frame metric $\tilde{g}_{\mu\nu}$ as an FLRW metric
\begin{eqnarray}
d\tilde{s}^2 = -d\tilde{t}^2 + \tilde{a}^2(\tilde{t})
d\bar{x}^2\,,
\label{nfrw}
\end{eqnarray}
by using the redefinitions $$d\tilde{t}=\sqrt{F(R)}\,\,dt\,,\,\,\,\,
\tilde{a}(t)=\sqrt{F(R)}\,\,a(t)\,.$$ The Hubble parameters of the
Jordan frame metric and the Einstein frame metric are now related by
the equation
\begin{eqnarray}
H=\sqrt{F}\left(\tilde{H}-\sqrt{\frac{\kappa}{6}}\,\phi'\right)\,,
\label{hhtilde}
\end{eqnarray}
where $\tilde{H}=\frac{\tilde{a}'(\tilde{t})}{\tilde{a}(\tilde t)}$, prime now stands
for $d/d\tilde{t}$. We can now write the Friedmann equations in the
Einstein frame
\begin{eqnarray}
\tilde{H}^2 &=& \frac{\kappa}{3}(\rho_{\phi} +
\tilde{\rho})\,,
\label{fridm1}
\\
\tilde{H}'&=& -\frac{\kappa}{2}[\phi'^2 +
(1+\omega)\tilde{\rho}]\,,
\label{hdot}
\end{eqnarray}
where
\begin{eqnarray}
\rho_{\phi} = \frac12 \phi'^2
+ V(\phi)\,.
\label{phieng}
\end{eqnarray}
The other relevant dynamical equations in Einstein frame are the
Klein-Gordon equation for the scalar field
\begin{eqnarray}
\phi''+ 3\tilde{H} \phi' +
\frac{dV}{d\phi}=\sqrt{\frac{\kappa}{6}}(1-3\omega)\tilde{\rho}\,,
\label{phiev}
\end{eqnarray}
and the continuity equation for the hydrodynamic matter component
\begin{eqnarray}
\tilde{\rho}' + \sqrt{\frac{\kappa}{6}}(1-3\omega)\tilde{\rho}\phi'
+3\tilde{H}\tilde{\rho}(1+\omega)=0\,. 
\label{rtildev}
\end{eqnarray}
It is seen from the above equations that the scalar field and
hydrodynamic matter components in the Einstein frame
satisfy coupled differential equations.

Before we end this section we want to point out that we expect physics
to be the same as observed from both the conformal frames. Apparently
the two conformal frames may not look the same but the features that
seem to be different in the two frames may be due to the fact that one
needs to transform also the units he/she uses, between one frame and the
other.
\section{Description of a bouncing scenario}
\label{dbscen}

In this section we describe the scenario of a cosmological bounce in
Jordan frame from the point of view of both the frames. A cosmological
bounce for the homogeneous and isotropic FLRW universe is defined
mathematically by the conditions
\begin{eqnarray}  
H_b=0\,,\,\,\,\, {\rm and}\,\,\,\, \dot{H}_b>0\,.
\label{bconds}
\end{eqnarray}
where the subscript $b$ on a time dependent quantity denotes the value
of the quantity at the time of the bounce. Using Eq.~(\ref{fried}), the
first bouncing condition in the Jordan frame becomes
\begin{eqnarray}
\rho_{b}+\frac{R_{b}f^{\prime}_{b}-f_{b}}{2\kappa}=0\,.
\label{hzero}
\end{eqnarray}
Throughout the article we will assume that the matter component
satisfies the conditions
\begin{eqnarray}
  \rho \ge 0\,,\,\,\,\,\rho + P \ge 0\,,
 \label{econd}
\end{eqnarray}
in the Jordan frame\footnote{We are not calling this condition as the
  weak energy condition as the energy conditions are generally stated
  in the Einstein frame.}.  If matter satisfies the above condition
then the second bouncing condition in the Jordan frame becomes
\begin{equation}
\dot{R}^{2}_{b}f^{\prime \prime \prime}_{b}+\ddot{R}_{b}f^{\prime
  \prime}_{b}<0\,.
\label{hdotp1}
\end{equation}
Solving the modified dynamical equations, Eq.~(\ref{2ndeqn}) and Eq.~(\ref{cont}),
in the Jordan frame, while keeping in mind the constraint,
Eq.~(\ref{fried}), requires three conditions $H(t=0)$,
$\dot{H}(t=0)$ and $\ddot{H}(t=0)$. If we choose $t=0$ to be the
bouncing moment, then the bouncing conditions require
\begin{eqnarray}  
H(0)=0\,,\,\,\,\, {\rm and}\,\,\,\, \dot{H}(0)>0\,.
\end{eqnarray}
However, $\ddot{H}(0)$ can be positive, negative or zero. If $\ddot{H}(0)=0$,
then we have a completely symmetric bounce, i.e., the contracting phase
of the bounce is a mirror image of the expanding phase of the
bounce. If $\ddot{H}(0)>0$ ($\ddot{H}(0)<0$) then the evolution of
the scale factor is steeper in the expansion (contraction) phase.

Let us now try to visualize how a Jordan frame bounce looks like from
the Einstein frame. Any cosmological evolution in the Einstein frame
is governed by the nature of the potential $V(\phi)$ and how the
scalar field moves on the potential. From the definition of the scalar
field $\phi$, we see that
\begin{eqnarray}
\phi_b=\sqrt{\frac{3}{2\kappa}}\ln f'(R_b)=\sqrt{\frac{3}{2\kappa}}\ln f'(6\dot{H}_b)\,,
\end{eqnarray}
and
\begin{eqnarray}
\left.\frac{d\phi}{d\tilde{t}}\right|_b=\sqrt{\frac{3}{2\kappa}}
\frac{\dot{R}_bf''(R_b)}{f'^{3/2}(R_b)}
=\sqrt{\frac{3}{2\kappa}}\frac{6\ddot{H}_bf''(6\dot{H}_b)}{f'^{3/2}(6\dot{H}_b)}\,.
\end{eqnarray}
Therefore the Jordan frame bouncing conditions applied on $\dot{H}_b$
and $\ddot{H}_b$ determines the Einstein frame values of $\phi_{b}$
and $\phi'_{b}$\footnote{We want to remind the reader at this point
  that a prime on $f$ implies a derivative with respect to $R$ where
  as a prime on $\phi$ implies a derivative with respect to
  $\tilde{t}$.} at $\tilde{t}=0$. Note that if the viability
conditions $f'>0$, $f''>0$ of an $f(R)$ gravity is respected, then the
sign of $\ddot{H}_b$ determines the sign of $\phi'_{b}$, which is in
turn related to the time symmetrical or asymmetrical nature of the
bounce.  The $H_b=0$ condition, from Eq.~(\ref{hhtilde}) implies
\begin{eqnarray}
\tilde{H}_b=\frac{\kappa}{6}\phi'_{b}.
\end{eqnarray}
Before we close this section we will like to present a short
discussion on the difficulties of attaining a bouncing solution in
$f(R)$ theory of gravity. One may work with $f(R)=A R^{1+\delta}$
where $A$ and $\delta$ are constants. For stability
$f^\prime=A(1+\delta)R^\delta >0$ and
$f^{\prime\prime}=A\delta(1+\delta)R^{\delta -1} >0$. From the
bouncing condition in the Jordan frame one can easily check that for a
successful bounce one requires $A\delta \le 0$ if we assume that $\rho
\ge 0$ for the hydrodynamic fluid. These conditions make
$f^{\prime\prime} <0$ near the bounce where we know that $R_b >0$ and
consequently $f(R)=A R^{1+\delta}$ cannot describe a gravitationally
stable bounce. In the quadratic model we have $f(R)=\lambda + R +
\alpha R^2$ where $\lambda\,,\,\alpha$ are constants and for a bounce
$\alpha<0$. In this case it is obvious that $f^{\prime\prime} <0$
throughout the bounce making the theory unstable. Even cubic gravity
models, where $f(R)=R + \alpha R^2 + \beta R^3$ where
$\alpha,\,,\beta$ are constants and for bounce $\alpha<0$ and
$\beta>0$, can accommodate cosmological bounces. One can tune the
parameters in such a way that $f^\prime>0$ for all values of $R$
\cite{Barrow:1988xh} by fixing the values of the parameters, but then
it is seen that there are two branches corresponding to positive and
negative values of $f^{\prime\prime}$ \cite{Bhattacharya:2015nda}. The
unstable branch remains a reality in such theories and the theory
becomes more complex as there appears a point where
$f^{\prime\prime}=0$ which is a singular point in $f(R)$ gravity. In
this way one can go on to show that even $f(R)=R+\alpha R^m$ where
$\alpha$ and $m$ are real constants and $m$ is an integer greater than
one gives rise to an unstable bounce. It was shown in
Ref.\cite{Bhattacharya:2015nda} that no polynomial $f(R)$ gravity can
simultaneously void both ghost and tachyonic instability for all $R$.
All these examples show that it is very difficult to find a $f(R)$
which gives rise to a stable cosmological bounce. In this regard we
show that the exponential $f(R)$ gravity theory, as chosen in the
present article, can produce perfectly stable cosmological
bounces. Our model is explained in the next section.
\section{Bounce in exponential gravity}
\label{begs}
\begin{figure}
\centering
\includegraphics[scale=.8]{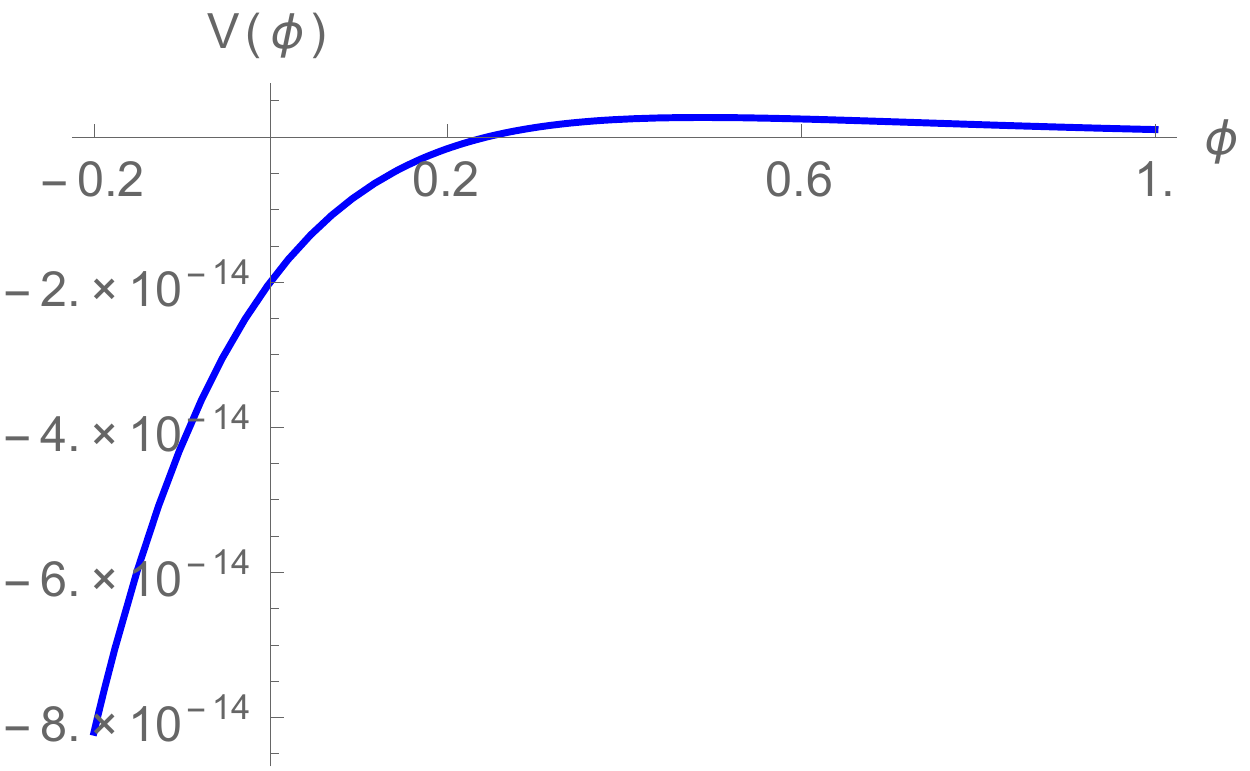}
\caption{Figure showing the nature of the scalar field potential in
  the Einstein frame where $f(R)=(1/\alpha)\exp(\alpha R)$. In the above figure
  the field and its potential both are expressed in Planck units
  (where the Planck mass is set as unity) in which
  $\alpha=10^{12}$.}
\label{pot_phi}
\end{figure}

In this section we will study cosmological bounce in $f(R)$ gravity
where the form of $f(R)$ is given by
\begin{eqnarray}
  f(R)=\frac{1}{\alpha}\exp(\alpha R)
\end{eqnarray}
where  $\alpha>0$. Note that for this gravity theory
\begin{eqnarray}
f'(R)=e^{\alpha R}>0\,,\,\,\,\, {\rm and}\,\,\,\, f''(R)=\alpha e^{\alpha R}>0\,,
\end{eqnarray}
implying that this theory is free from both ghost instability and
tachyonic instability for all values of $R$. Moreover,
\begin{eqnarray}
Rf'-f=\left(R-\frac{1}{\alpha}\right)e^{\alpha R}\,,
\end{eqnarray}
so we can see from the bouncing condition in Eq.~(\ref{hzero}) that in this
case both matter and matter less bounce is possible. For 
bounce in presence of matter, 
\begin{eqnarray}
R_b <\frac{1}{\alpha}\,,
\end{eqnarray}
whereas for a matter less bounce
\begin{eqnarray}
R_b=\frac{1}{\alpha}\,.
\end{eqnarray}
The cosmological constant term plays a crucial role in producing the
bounce If we want to remove the cosmological constant term by taking a
theory like $f(R)=(1/\alpha)[\exp(\alpha R)-1]$ with positive
$\alpha$, then we have
\begin{eqnarray}
Rf'-f=\frac{1}{\alpha}[(\alpha R-1)e^{\alpha R}+1]\,,
\end{eqnarray}
which is always positive for any positive value of $R$. This implies
that this theory of gravity cannot support a cosmological bounce in
either matter or vacuum background. It is precisely the cosmological
constant term that helps to achieve a bounce.

The Einstein frame scalar field for exponential
gravity is
\begin{eqnarray}
\phi=\sqrt{\frac{3}{2\kappa}}\ln f'=\sqrt{\frac{3}{2\kappa}}\alpha R\,,
\end{eqnarray}
whose potential in the Einstein frame is given by
\begin{eqnarray}
V(\phi)=\frac{1}{2\kappa\alpha}\left(\sqrt{\frac{2\kappa}{3}}\phi -1\right)
e^{-\sqrt{2\kappa/3}\phi}\,.
\end{eqnarray}
The shape of the above potential is shown in Fig.~\ref{pot_phi}. The
value of $\alpha=10^{12}$ in our system of units where all the
dimensional parameters are ultimately expressed in units of the Planck
mass $M_P$. For convenience we assume $M_P=1$ throughout the article.
The potential crosses the $V(\phi)$ axis at $\phi=0$ and the $\phi$ axis
at $\phi=\sqrt{\frac{3}{2\kappa}}$. As the gravitational part of the
theory, describing the bounce, becomes essentially GR in the Einstein
frame, it is easier and interesting to start from the Einstein frame
description and track the bounce as the movement of the scalar field
on the scalar potential. Firstly, note that the values $R=0$ and
$R=1/\alpha$ in the Jordan frame corresponds to the values $\phi=0$
and $\phi=\sqrt{3/(2\kappa)}$ in the Einstein frame. Since $R_b$ has a
limited range for bounce in presence of matter where the range is
$0<R_b<1/\alpha$ in the Jordan frame, $\phi_b$ is limited in the range
$0<\phi_b<\sqrt{3/(2\kappa)}$ in the Einstein frame. This means that
if we want to impose the Einstein frame intermediate\footnote{We call
  the conditions as intermediate instead of initial conditions. The
  reason being that we impose our conditions on the dynamical system
  at $t=\tilde{t}=0$ and look at the system at both negative and
  positive times.} conditions at the time of bounce in the Jordan
frame ($t=\tilde{t}=0$), then the scalar field in the Einstein frame
can start from only a very small part of the curve in
Fig.\ref{pot_phi}, which is in the fourth quadrant.

\begin{figure}[t!]
\begin{minipage}[b]{0.5\linewidth}
\centering
\includegraphics[scale=.5]{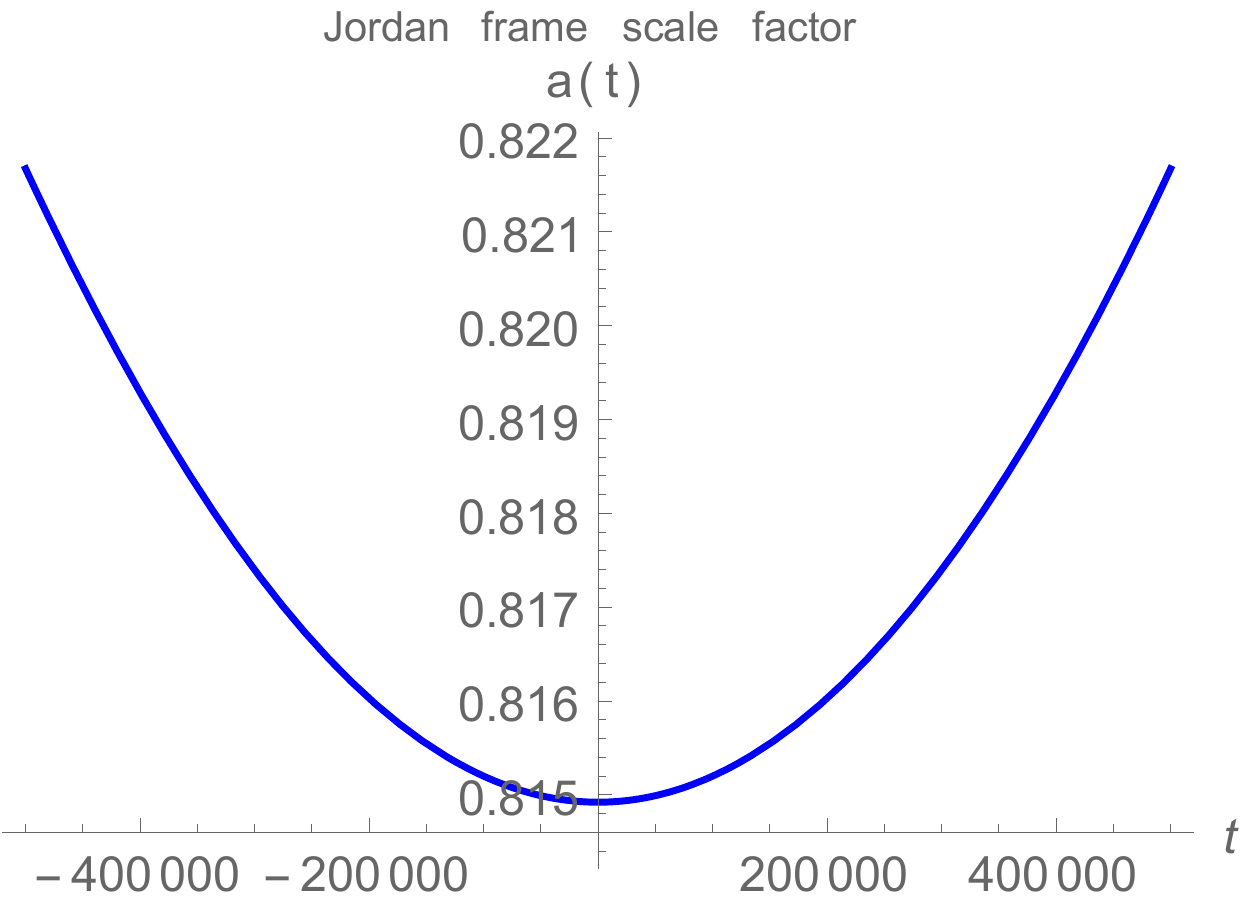}
\caption{Time evolution of the scale factor in the Jordan frame for a
  bounce in radiation background in exponential gravity found with the
  initial conditions $\phi(0)=0.1$, $\phi'(0)=0$,
  $\tilde{H}(0)=\sqrt{\kappa/6}\phi'(0)$.}
\label{arad1}
\end{minipage}
\hspace{0.2cm}
\begin{minipage}[b]{0.5\linewidth}
\centering
\includegraphics[scale=.5]{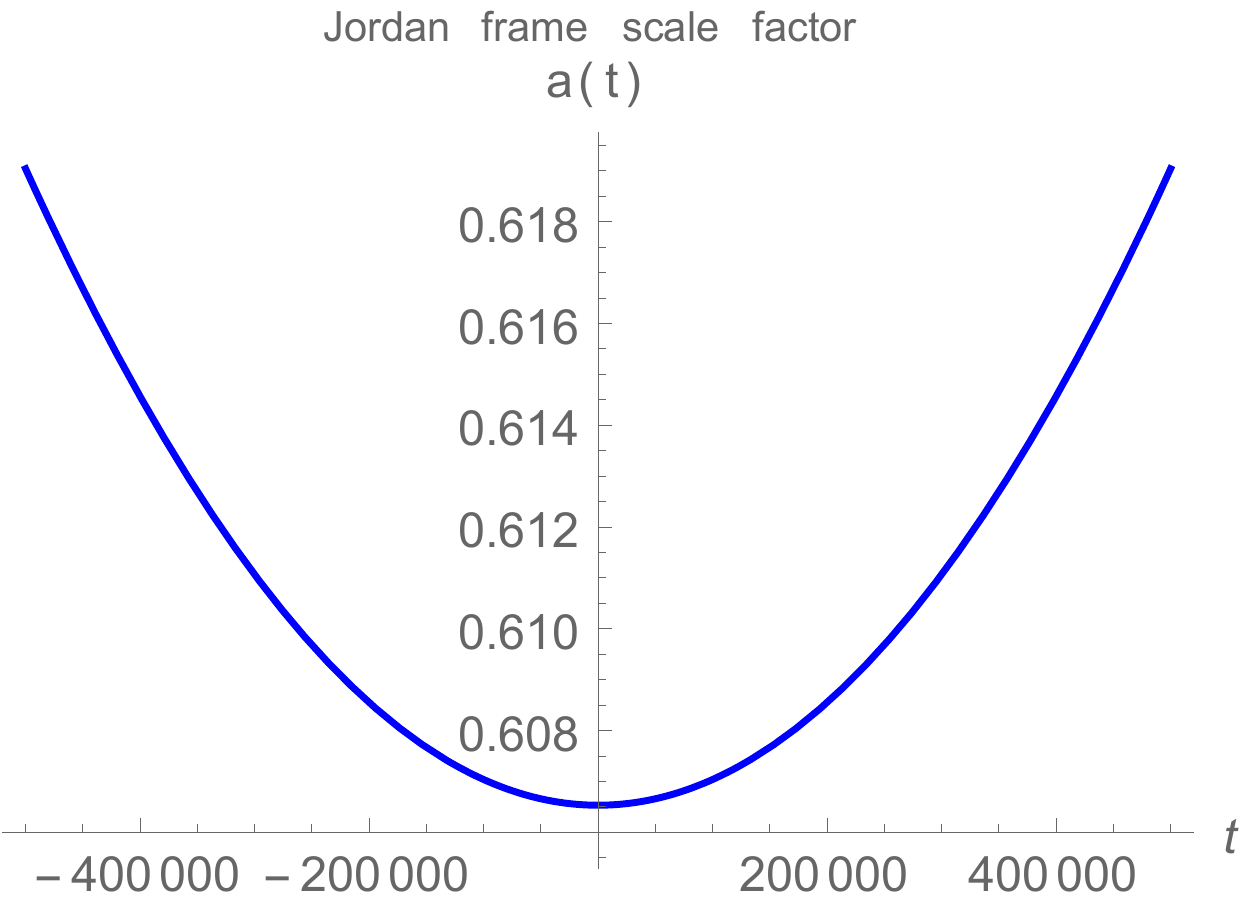}
\caption{Time evolution of the scale factor in the Jordan frame for a
  bounce in vacuum background in exponential gravity found with the
  initial conditions $\phi(0)=\sqrt{3/(2\kappa)}$, $\phi'(0)=0$,
  $\tilde{H}(0)=\sqrt{\kappa/6}\phi'(0)$.}
\label{avac1}
\end{minipage}
\end{figure}

We may recall from our discussion of the previous section that the
sign of $\phi'_b$, i.e., whether the field is rolling up or down the
potential when the bounce has happened is related to whether the time
evolution of the scale factor is completely symmetric or steeper on
one side and flatter on the other. The dependence of the nature of the
time evolution on the initial conditions at bounce is elaborated by
Figs.\ref{arad1}, \ref{avac1}. 
In both the figures we show a symmetric bounce, where the intermediate
conditions are specified on the figure captions. As exponential
gravity can accommodate bounces in presence of hydrodynamic matter and
vacuum the two figures show two different kind of universes, one filled
with radiation and the other devoid of matter. For symmetric bounces
we wee that $\phi^\prime(0)=0$ which translates to $\ddot{H}(0)=0$ in the
Jordan frame. For purely symmetrical bounces one must have
$\ddot{H}(0)=H(0)=0$ and  $\dot{H}(0)>0$ in the Jordan frame. The last statement
is true for all kinds of bounces in metric $f(R)$ theory. 

\begin{figure}[t!]
\begin{minipage}[b]{0.5\linewidth}
\centering
\includegraphics[scale=.5]{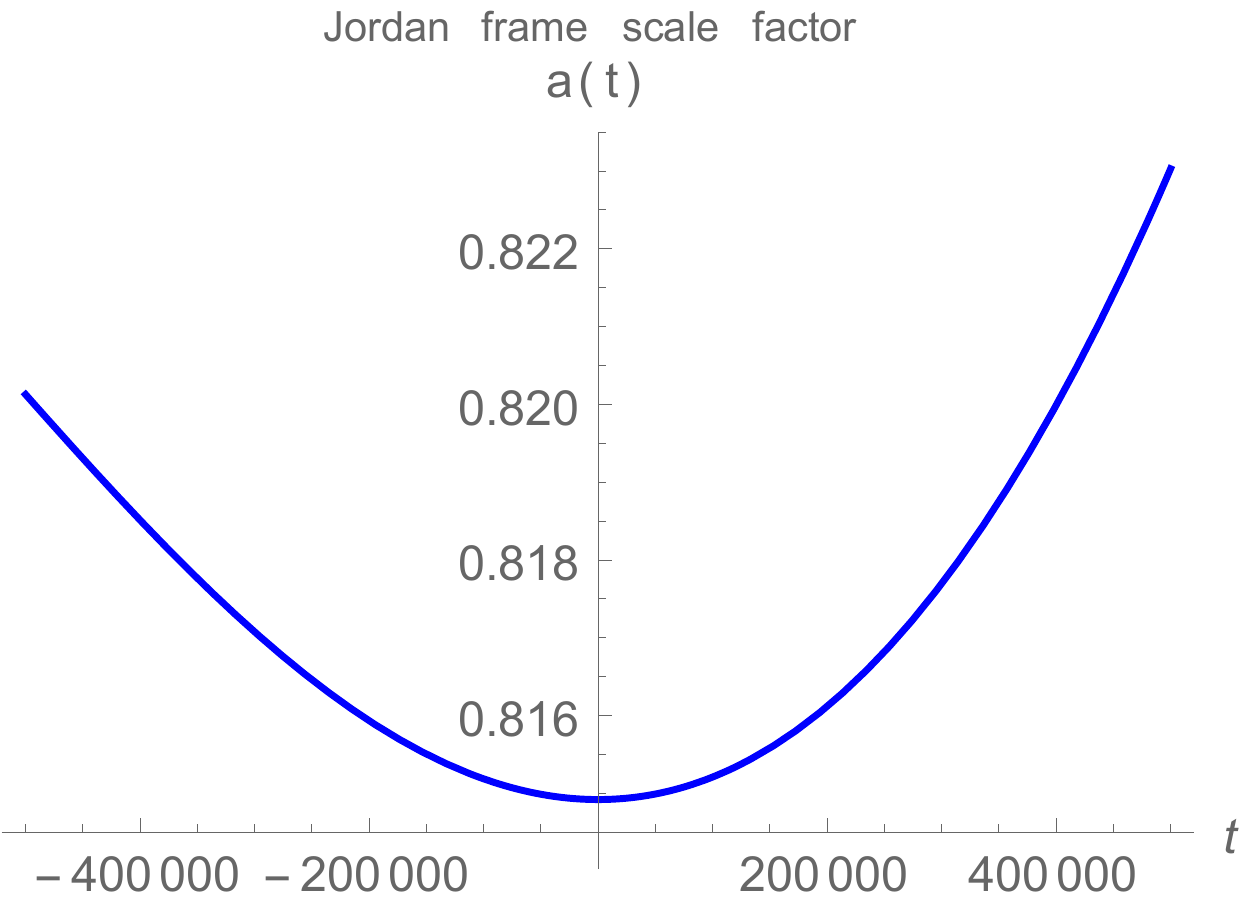}
\caption{Time evolution of the scale factor in the Jordan frame for a
  bounce in radiation background in exponential gravity found with the
  initial conditions $\phi(0)=0.1$, $\phi'(0)=10^{-7}$,
  $\tilde{H}(0)=\sqrt{\kappa/6}\phi'(0)$.}
\label{arad3}
\end{minipage}
\hspace{0.2cm}
\begin{minipage}[b]{0.5\linewidth}
\centering
\includegraphics[scale=.5]{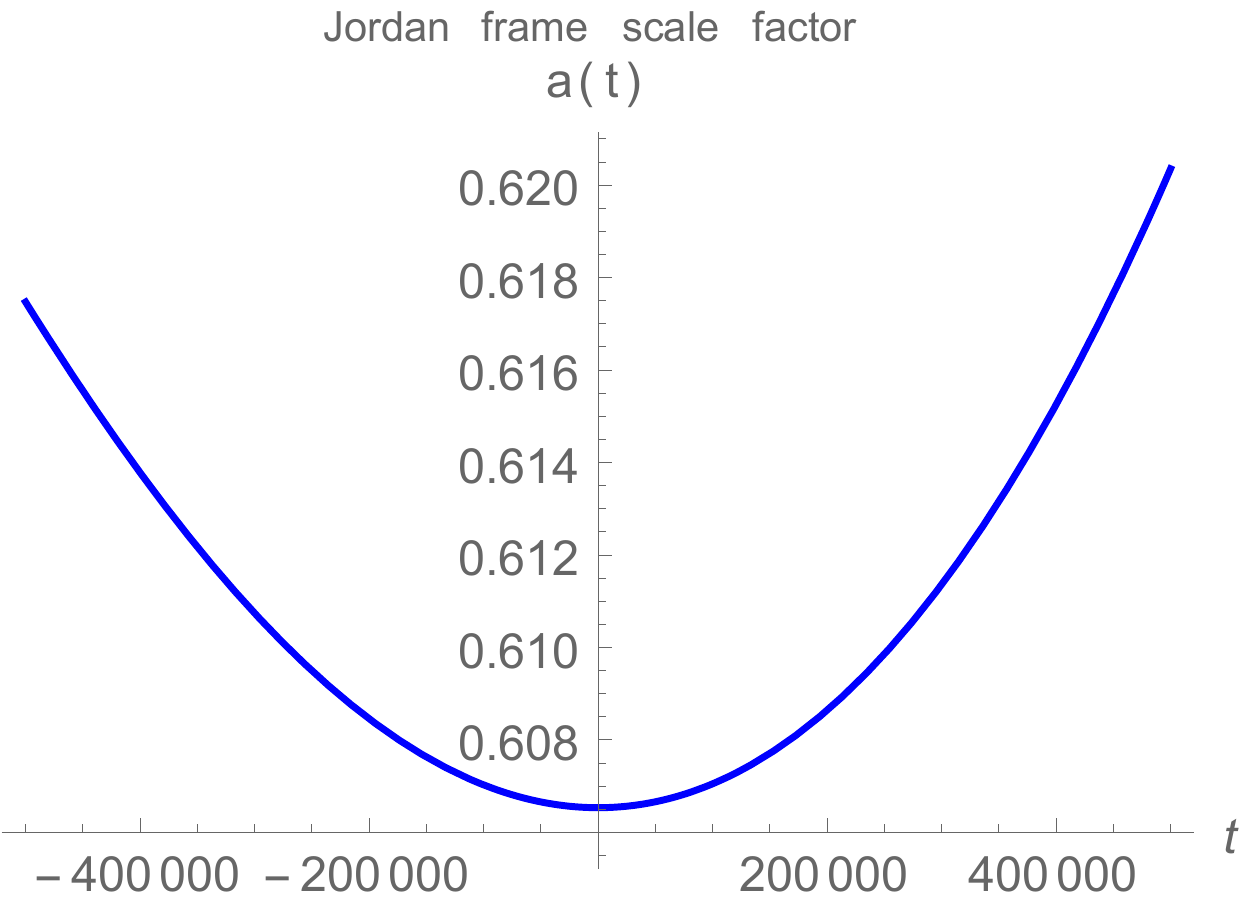}
\caption{Time evolution of the scale factor in the Jordan frame for a
  bounce in vacuum background in exponential gravity found with the
  initial conditions $\phi(0)=\sqrt{3/(2\kappa)}$,
  $\phi'(0)=10^{-7}$, $\tilde{H}(0)=\sqrt{\kappa/6}\phi'(0)$.}
\label{avac3}
\end{minipage}
\end{figure}

One can also model asymmetric bounces in $f(R)$ theories. In our case
we show two asymmetric bounces in Fig.~\ref{arad3} and
Fig.~\ref{avac3}. The bounce depicted in Fig.~\ref{arad3} is assisted
by radiation where as the other bounce takes place in vacuum. In the
case of bounce in presence of radiation one can easily calculate from
the intermediate conditions that $\ddot{H}(0)=(2/(3e^2))\times
10^{-19}$ and for vacuum bounce one gets
$\ddot{H}(0)=(2\sqrt{e}/3)\times 10^{-19}$\,. In these cases one has
non-zero values of the second time derivative of the Hubble parameter
in the Jordan frame. In all of the bounces we have discussed in this
section $\phi$ in the Einstein frame remains positive in sign which
implies that $R>0$ for all the bounces, in the Jordan frame. In the
effective theory approach exponential gravity becomes similar to GR
near small $R$ values when one can neglect the higher powers (starting
from the quadratic one) of $R$ in the exponential $f(R)$.  The theory
presented in this paper is consistent when we use the theory for
positive non-zero values of $R$ in the Jordan frame. During the end
stages of the bouncing scenario, in all the cases, $R$ tends to zero
by which one is very near the GR limit. In the bouncing scenarios
presented in this article we do not specify how the effective $f(R)$
theory can transform to GR near $R \sim 0$, we hope some new physics
is involved during this phase. 

On the other hand if we do not take the $f(R)$ theory presented in our
article as some form of effective theory, which can change cosmology
only for high values of the Ricci scalar, then something interesting
happens.  If the bounce is not due to some effective change in the
gravitational part of the Lagrangian, and the $f(R)$ modifications are
valid for all values of $R$ including $R=0$ then there are new
possible dynamical configurations of the universe. These new
configurations depend upon the intermediate conditions applied to
produce the dynamics of the bounce.
\section{Evolution of metric perturbations through the bounce}
\label{ptb}

In this section we discuss the evolution of scalar cosmological
perturbation through a bounce in $f(R)$ gravity. We also briefly opine
on tensor perturbations in bouncing cosmologies guided by $f(R)$
theories before we end this section. We express the scalar perturbation
equations in terms of conformal time defined as follows,
\begin{equation}
d\eta=\frac{dt}{a(t)}=\frac{d\tilde{t}}{\tilde{a}(\tilde{t})}.
\label{conf_time}
\end{equation}
In the domain of linear perturbations, the scalar perturbed FLRW
metric has two gauge invariant degrees of freedom. In the longitudinal
gauge this can be expressed as,
\begin{eqnarray}
ds^{2}=a^{2}(\eta)\left[ -(1+2\Phi)d\eta^{2}+(1-2\Psi)\delta_{ij} 
dx^{i}dx^{j}\right].
\label{ptbd_flrw_metric}
\end{eqnarray}
Here $\Phi$ and $\Psi$ are the two gauge invariant perturbation
degrees of freedom, also called the Bardeen potentials. If only a
single barotropic matter component is present, the perturbation in the
matter sector can be assumed to be adiabatic, so that the sound
velocity can be defined as
\begin{equation}
c_s^2=\frac{\delta p}{\delta\rho}=\omega,
\label{sound_velocity}
\end{equation}
where $\omega$ is the constant equation of state of the barotropic
matter component. In this section we will be working with conformal
time and a derivative with the conformal time will be represented by
a prime above. Consequently we will specify the derivatives of $f(R)$
with respect to $R$ as $f_R$ and higher derivatives with respect to
$R$ as $f_{RR}$ and so on\footnote{The new conventions which are at
  odds with our previous convention becomes necessary as conformal time
  is involved in the discussions. From the next section we will use
  the old conventions.}. The $00$, $ii$, $0i$, $ij-th(i\neq j)$
components of the linearized $f(R)$ field equations in the Fourier
space are\cite{Matsumoto:2013sba}
\begin{eqnarray}
  f_R[-k^2(\Phi+\Psi)-3\mathcal{H}(\Phi^\prime+\Psi^\prime)+(3\mathcal{H}^\prime -6\mathcal{H}^2)
    \Phi-3\mathcal{H}^\prime\Psi]+f_R^\prime(-9\mathcal{H}\Phi+3\mathcal{H}\Psi-3\Psi^\prime)=
  \kappa a^2\rho\delta, &
\label{ptbd_00}
\\
f_R[\Phi^{\prime\prime}+\Psi^{\prime\prime}+3\mathcal{H}(\Phi^\prime+\Psi^\prime)+3\mathcal{H}^\prime\Phi+
  (\mathcal{H}^\prime+2\mathcal{H}^2)\Psi] +f_R^\prime(3\mathcal{H}\Phi-\mathcal{H}\Psi+3\Phi^\prime)
+f_R^{\prime\prime}(3\Phi-\Psi)=\kappa a^2c_s^2\rho\delta, & 
\label{ptbd_ii}
\\
f_R[\Phi^\prime+\Psi^\prime+\mathcal{H}(\Phi+\Psi)]+f_R^\prime(2\Phi-\Psi)=
-\kappa a^2\rho(1+\omega)v, &
\label{ptbd_i0}
\\
\Phi-\Psi-\frac{2f_{RR}}{a^2f_R}[3\Psi^{\prime\prime}+6(\mathcal{H}^\prime+\mathcal{H}^2)\Phi
  +3\mathcal{H}(\Phi^\prime+3\Psi^\prime)-k^2(\Phi-2\Psi)]=0. & 
\label{ptbd_ij}
\end{eqnarray}
Here we have defined the matter density perturbation as
$\delta=\frac{\delta\rho}{\rho}$ and the perturbed velocity potential
as $\delta u_i=\frac{1}{a}\partial_iv$. There are a total of four
dynamical scalar perturbation quantities, namely, $\Phi$, $\Psi$,
$\delta$, $v$, and two constraint equations between them, namely,
Eq.~(\ref{ptbd_00}) and Eq.~(\ref{ptbd_i0}). Therefore only two of them are
independent. For our convenience, we can take them to be the two
metric degrees of freedom $\Phi$, $\Psi$.
\begin{figure}[t!]
\begin{minipage}[b]{0.5\linewidth}
\centering
\includegraphics[scale=.5]{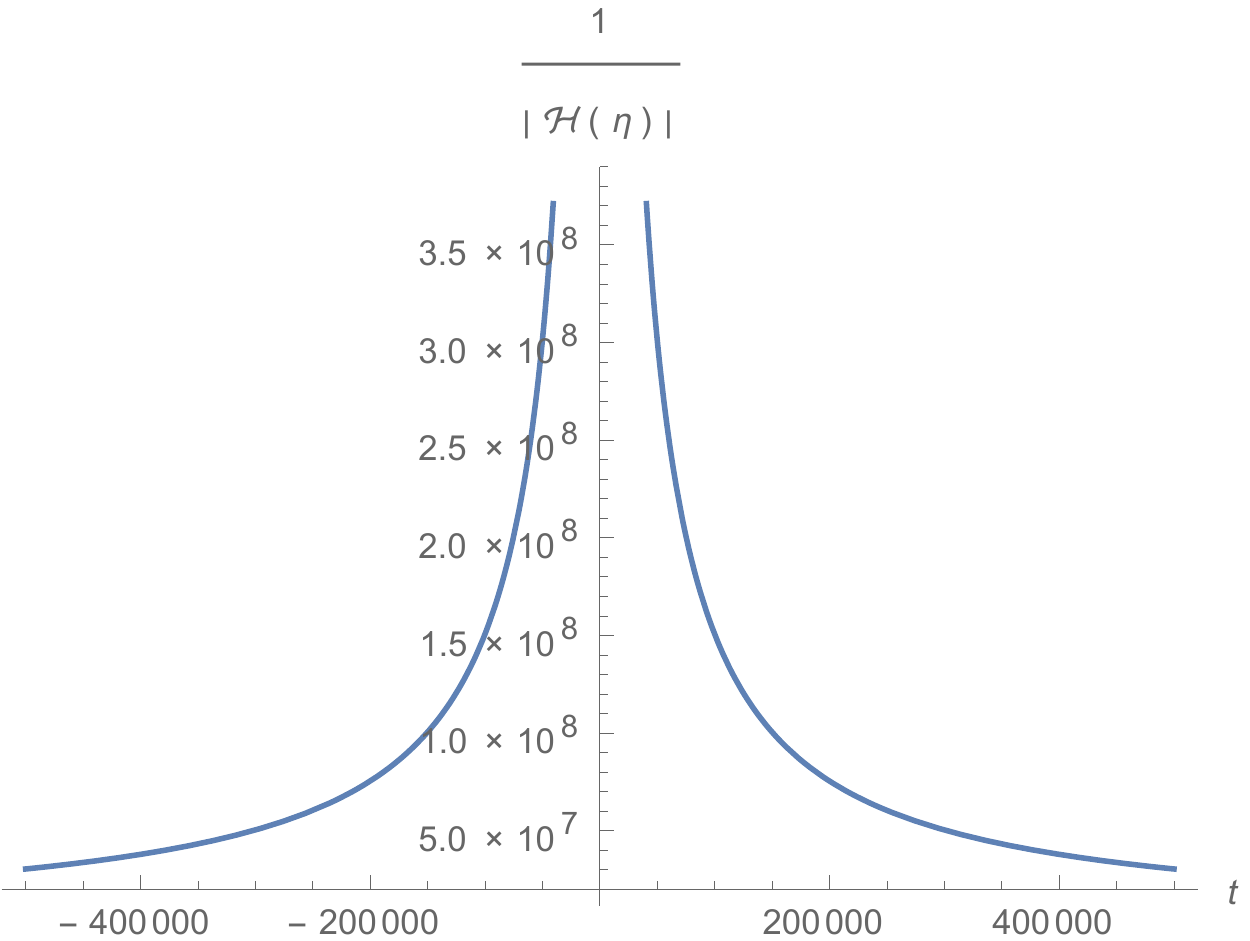}
\caption{Time evolution of the comoving Hubble radius for the time-symmetric
  bouncing solution of Fig.\ref{arad1} in radiation background.} 
\label{hubradius_rad}
\end{minipage}
\hspace{0.2cm}
\begin{minipage}[b]{0.5\linewidth}
\centering
\includegraphics[scale=.5]{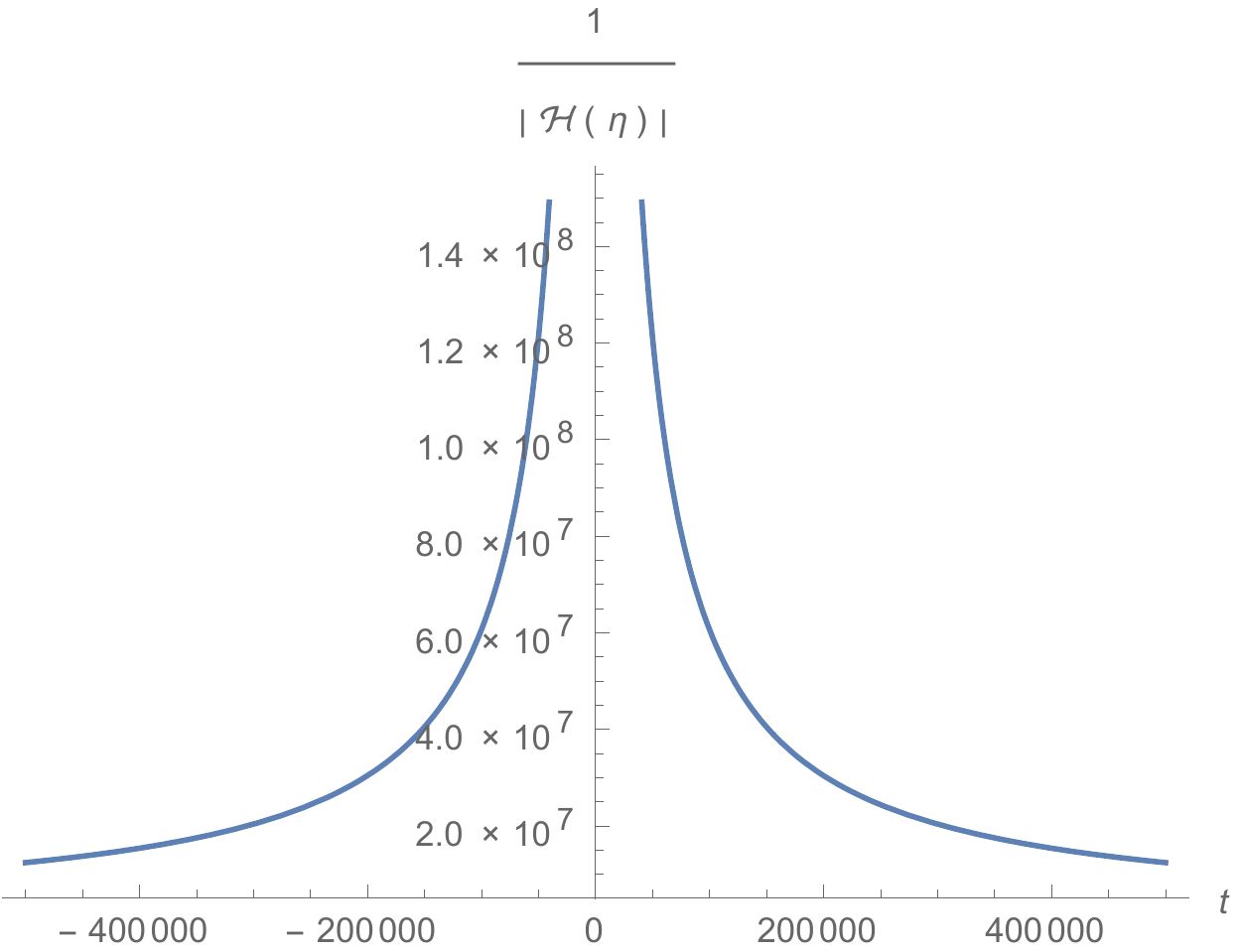}
\caption{Time evolution of the comoving Hubble radius for the time-symmetric bouncing
  solution of Fig.\ref{avac1} in vacuum background.}
\label{hubradius_vac}
\end{minipage}
\end{figure} 

From Eqs.~(\ref{ptbd_00}) and (\ref{ptbd_ii}), we can eliminate $\delta$ to write,
\begin{eqnarray}
  f_R[\Phi^{\prime\prime}+\Psi^{\prime\prime}+3\mathcal{H}(1+c_s^2)(\Phi^\prime+\Psi^\prime)+
    c_s^2(k^2+6\mathcal{H}^2)\Phi+3\mathcal{H}^\prime(1-c_s^2)\Phi+\mathcal{H}^\prime(1+3c_s^2)
    \Psi\nonumber\\
    +(2\mathcal{H}^2+c_s^2k^2)\Psi]+f_R^\prime[\mathcal{H}(1+3c_s^2)(3\Phi-\Psi)+
    3\Phi^\prime+3c_s^2\Psi^\prime]+f_R^{\prime\prime}(3\Phi-\Psi)=0.
\label{ptbd_ii00}
\end{eqnarray}
The equations Eq.~(\ref{ptbd_ij}) and Eq.~(\ref{ptbd_ii00}) can be
solved to get the solutions $\Phi(\eta)$ and $\Psi(\eta)$.  Once
$\Phi$, $\Psi$ is known, $\delta$ and $v$ is determined from
Eq.~(\ref{ptbd_00}) and Eq.~(\ref{ptbd_i0}). We calculate the gauge
invariant comoving curvature perturbation as
\begin{equation}
\mathcal{R}=\Psi+\frac{\mathcal{H}\delta}{1+\omega}.
\end{equation}

For the case of vacuum the right hand sides of Eqs.~(\ref{ptbd_00},
\ref{ptbd_ii}, \ref{ptbd_i0}) all vanish. In this case we can just
solve the two coupled second order differential equations
Eq.~(\ref{ptbd_ii}) and Eq.~(\ref{ptbd_ij}) to obtain $\Phi(\eta)$ and
$\Psi(\eta)$. In this case the comoving curvature perturbation is just
\begin{equation}
\mathcal{R}=\Psi.
\end{equation}
\subsection{Scalar perturbation evolution through bounce}

In this section we present the numerical solution of the perturbation
equations by assuming suitable initial conditions. Let us first
discuss some general conclusions regarding the forms of the solutions
of the perturbation equations in the context of a nonsingular bounce.
First of all note that the Hubble radius diverges at the bounce (see
Figs.\ref{hubradius_rad} and \ref{hubradius_vac}). Therefore as the
bounce approaches, all the perturbation modes become sub-Hubble near
the bounce. We consider perturbation modes with wavenumber $k=10^{-8}$ (in Planck units)
to illustrate our point regarding perturbation growth near the bounce point.

In general for all models of non-singular bouncing cosmologies, near
the bounce point when $\eta \sim 0$, the scale-factor can be well
approximated by an even function of the conformal time. In this
section we will be working in the vicinity of the $\eta=0$ point and
consequently assume the scale-factor to be an an even function.  As
the scale-factor is an even function of time, the Hubble parameter
would be an odd function of conformal time, as it is first order
derivative (with respect to $\eta$) of the former. Likewise, $R(\eta)$
and all the functions of it, like $f_R$ , $f_{RR}$ etc. would be even
functions of time.

Below we constrain ourselves to the case of a bouncing solution in
radiation background, although our considerations hold true also for a
bounce in vacuum background. Both the perturbation evolution equations
(Eqs.~\ref{ptbd_ij} and \ref{ptbd_ii00}) that we use to solve for
$\Phi(\eta)$ and $\Psi(\eta)$ are of the form
\begin{equation}
\left[\phi_2(\eta)D_{\eta}^2+\phi_1(\eta)D_{\eta}+\phi_0(\eta)\right]\Phi(\eta)
+\left[\psi_2(\eta)D_{\eta}^2+\psi_1(\eta)D_{\eta}+\psi_0(\eta)\right]\Psi(\eta)=0
\label{ptbd_generic}
\end{equation}
where $\phi_0(\eta)$, $\phi_1(\eta)$, $\phi_2(\eta)$, $\psi_0(\eta)$,
$\psi_1(\eta)$, $\psi_2(\eta)$ are functions defined by the background
evolution and $D_{\eta}\equiv\frac{d}{d\eta}$. In case of a completely
time-symmetric evolution near $\eta=0$, $\phi_0(\eta)$,
$\phi_2(\eta)$, $\psi_0(\eta)$, $\psi_2(\eta)$ are even functions
whereas $\phi_1(\eta)$, $\psi_1(\eta)$ are odd functions. For example,
for Eq.~(\ref{ptbd_ij}) we have
\begin{figure}[t!]
\begin{minipage}[b]{0.5\linewidth}
\centering
\includegraphics[scale=.5]{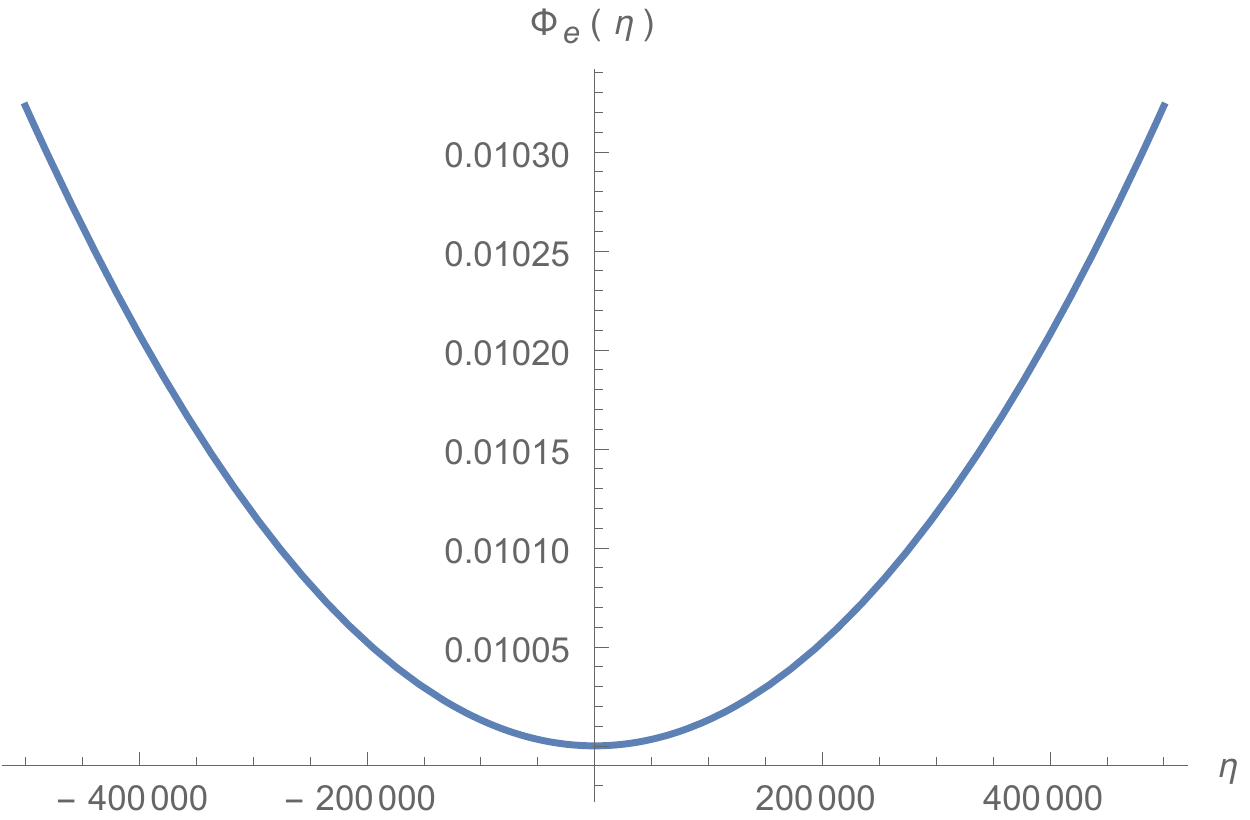}
\caption{Time evolution of $\Phi_e$ in the Jordan frame for the
  initial conditions $\Phi_e(0)=10^{-2}$, $\Psi_e(0)=-10^{-3}$,
  $\Phi_e'(0)=0$, $\Psi_e'(0)=0$. The background evolution is the one
  obtained by assuming the same initial conditions as mentioned in the
  caption of Fig.\ref{arad1}.}
\label{Phi1}
\end{minipage}
\hspace{0.2cm}
\begin{minipage}[b]{0.5\linewidth}
\centering
\includegraphics[scale=.5]{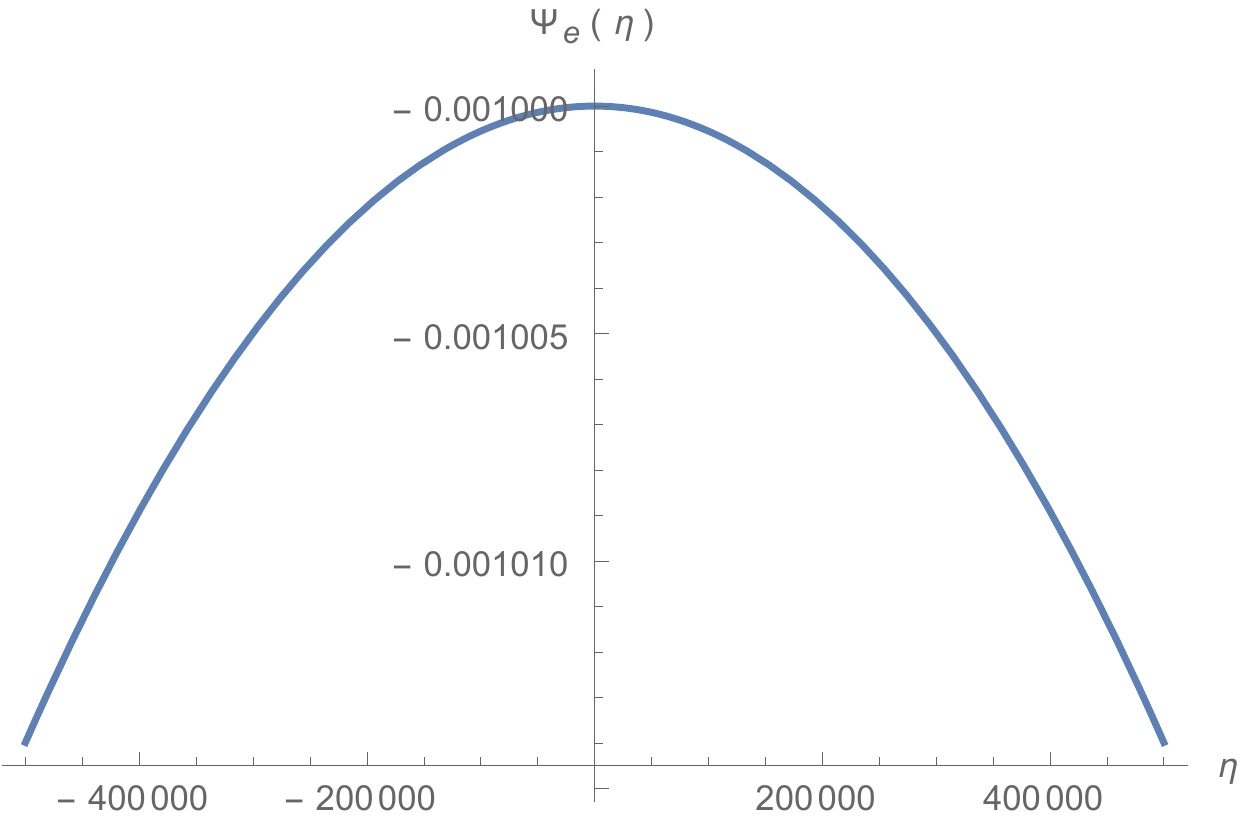}
\caption{Time evolution of $\Psi_e$ in the Jordan frame for the
  initial conditions $\Phi_e(0)=10^{-2}$, $\Psi_e(0)=-10^{-3}$,
  $\Phi_e'(0)=0$, $\Psi_e'(0)=0$. The background evolution is the one
  obtained by assuming the same initial conditions as mentioned in the
  caption of Fig.\ref{arad1}.}
\label{Psi1}
\end{minipage}
\end{figure} 
\begin{eqnarray}
\phi_2(\eta)&=& 0,\\
\phi_1(\eta)&=& -\frac{6 \mathcal{H}f_{RR}}{a^2 f_{R}},\\
\phi_0(\eta)&=& 1-\frac{12 f_{RR}(\mathcal{H}' +\mathcal{H}^2)}{a^2 f_{R}}+
\frac{2 k^2 f_{RR}}{a^2 f_{R}},\\
\psi_2(\eta)&=& -\frac{6 f_{RR}}{a^2 f_{R}},\\
\psi_1(\eta)&=& -\frac{18 \mathcal{H}f_{RR}}{a^2 f_{R}},\\
\psi_0(\eta)&=& -1- \frac{4 k^2 f_{RR}}{a^2 f_{R}} .
\end{eqnarray}
whereas for Eq.~(\ref{ptbd_ii00}) we have
\begin{eqnarray}
\phi_2(\eta)&=& f_{R},\\
\phi_1(\eta)&=& 3 \mathcal{H}f_{R} (1+c_{s}^2)+3R'f_{RR},\\
\phi_0(\eta)&=& f_{R}[c_{s}^2 (k^2-3( \mathcal{H}'-2 \mathcal{H}^2)) + 3 \mathcal{H}']+
3[ \mathcal{H} (1+3c_{s}^2) R' +R'']f_{RR}+3 R'^2 f_{RRR},\\
\psi_2(\eta)&=& f_{R},\\
\psi_1(\eta)&=& 3 \mathcal{H}f_{R} (1+c_{s}^2)+ 3c_{s}^2 R'f_{RR},\\
\psi_0(\eta)&=& f_{R}[c_{s}^2 (k^2+ \mathcal{H}') +  \mathcal{H}' +2 \mathcal{H}^2]-
    [\mathcal{H} (1+3c_{s}^2) R' +R'']f_{RR}- R'^2 f_{RRR}.
\end{eqnarray}
The above coefficient functions are seen to be non-singular near
$\eta=0$ showing that the perturbation evolution equation, as given in
Eq.~(\ref{ptbd_generic}), do not produce singular solutions near the
bounce point. Since $D_{\eta}\rightarrow-D_{\eta}$ when
$\eta\rightarrow-\eta$, the perturbation equations are time-symmetric,
and have even solutions $\Phi_e(\eta)$, $\Psi_e(\eta)$ and odd
solutions $\Phi_o(\eta)$, $\Psi_o(\eta)$. Note that, since the
perturbation evolution equations are linear differential equations, in
general any linear combination of the even and odd solutions
$c_1\Phi_e+c_2\Phi_o$ and $c_1\Psi_e+c_2\Psi_o$ is also a solution of
the perturbation evolution equations. The odd solutions
$\Phi_o(\eta)$, $\Psi_o(\eta)$ vanish at bounce, meaning that if these
solutions start with an absolute value less than unity prior to the
bounce, then it remains less than unity throughout the bouncing
phase. In other words, these odd solutions, if they start at a
perturbative level before the bounce, they remain at the perturbative
level throughout the bounce.  On the contrary the even solutions
$\Phi_e(\eta)$, $\Psi_e(\eta)$ must have a local extremum at
bounce.

\begin{figure}[t!]
\begin{minipage}[b]{0.5\linewidth}
\centering
\includegraphics[scale=.5]{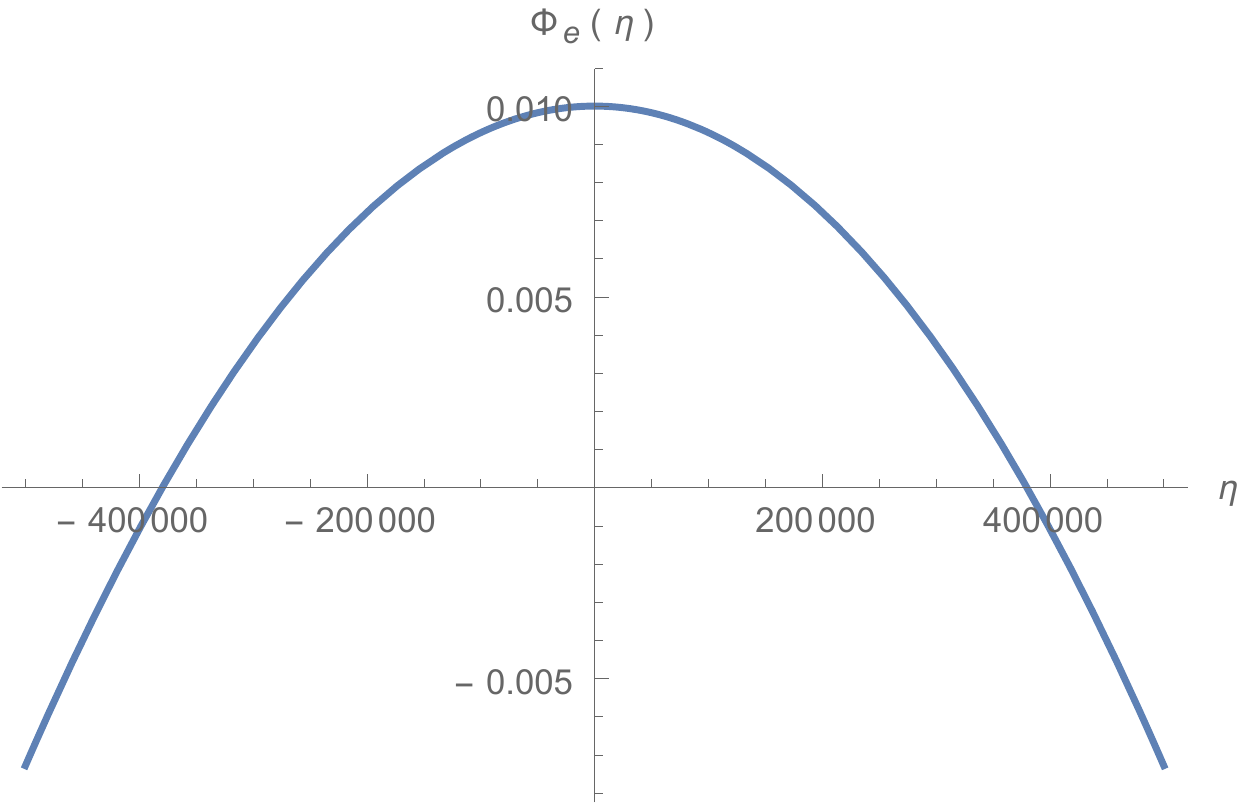}
\caption{Time evolution of $\Phi_e$ in the Jordan frame for the
  initial conditions $\Phi_e(0)=10^{-2}$, $\Psi_e(0)=1$,
  $\Phi_e'(0)=0$, $\Psi_e'(0)=0$. The background evolution is the one
  obtained by assuming the same initial conditions as mentioned in the
  caption of Fig.\ref{arad1}.}
\label{Phi3}
\end{minipage}
\hspace{0.2cm}
\begin{minipage}[b]{0.5\linewidth}
\centering
\includegraphics[scale=.5]{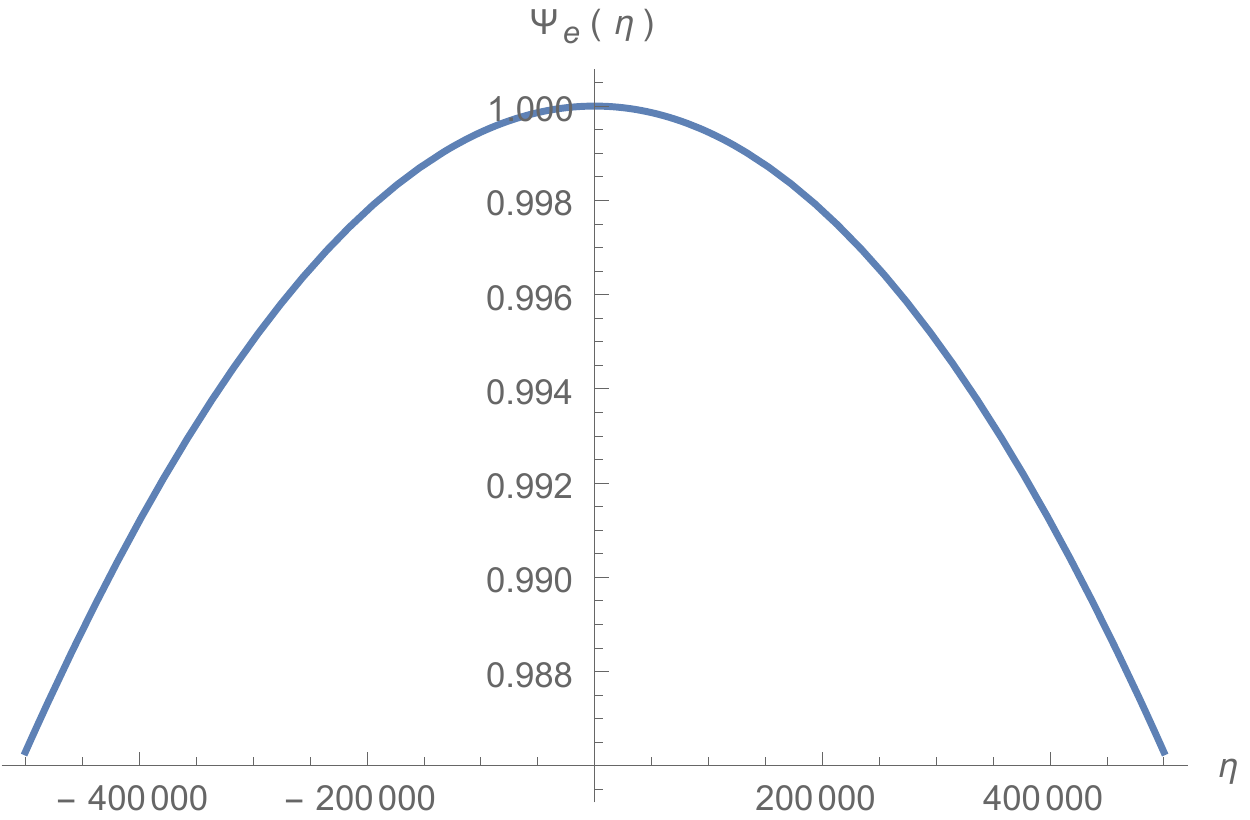}
\caption{Time evolution of $\Psi_e$ in the Jordan frame for the
  initial conditions $\Phi_e(0)=10^{-2}$, $\Psi_e(0)=1$,
  $\Phi_e'(0)=0$, $\Psi_e'(0)=0$. The background evolution is the one
  obtained by assuming the same initial conditions as mentioned in the
  caption of Fig.\ref{arad1}.}
\label{Psi3}
\end{minipage}
\end{figure}

\begin{figure}[t!]
\begin{minipage}[b]{0.5\linewidth}
\centering
\includegraphics[scale=.5]{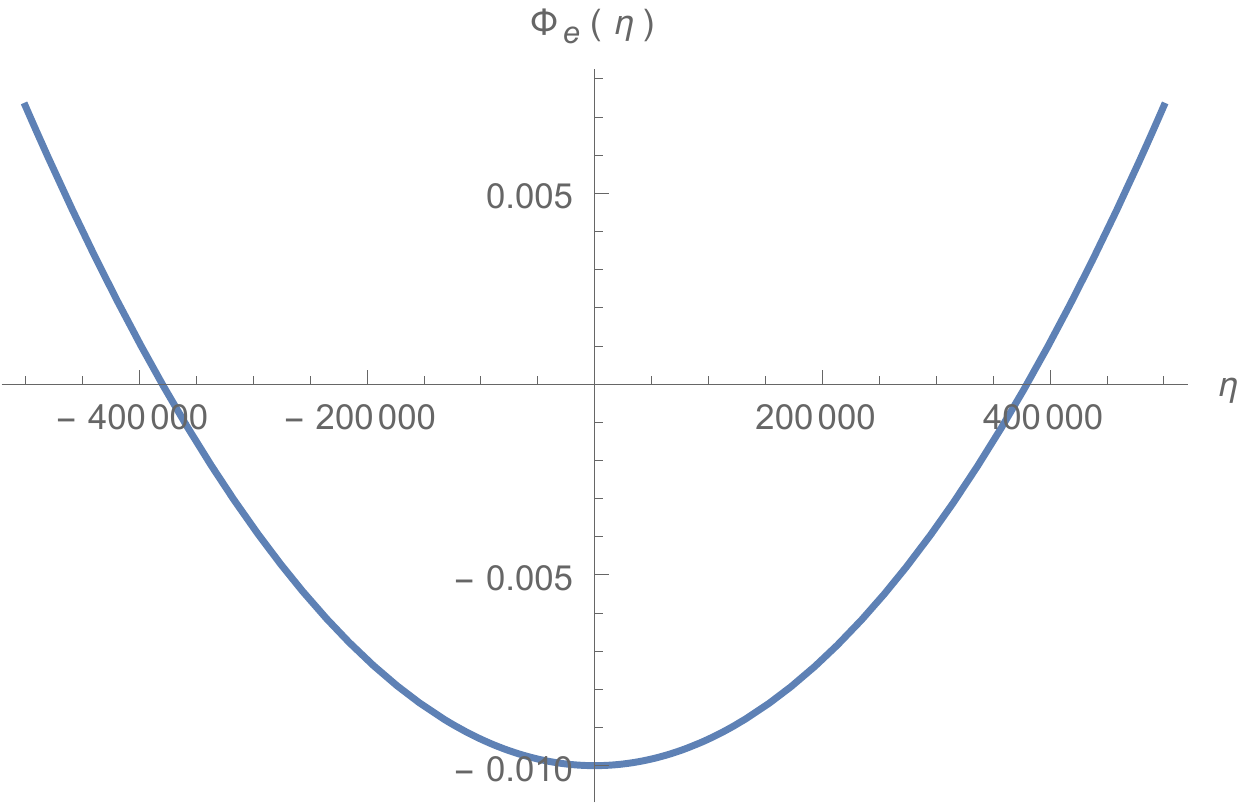}
\caption{Time evolution of $\Phi_e$ in the Jordan frame for the
  initial conditions $\Phi_e(0)=-10^{-2}$, $\Psi_e(0)=-1$,
  $\Phi_e'(0)=0$, $\Psi_e'(0)=0$. The background evolution is the one
  obtained by assuming the same initial conditions as mentioned in the
  caption of Fig.\ref{arad1}.}
\label{Phi4}
\end{minipage}
\hspace{0.2cm}
\begin{minipage}[b]{0.5\linewidth}
\centering
\includegraphics[scale=.5]{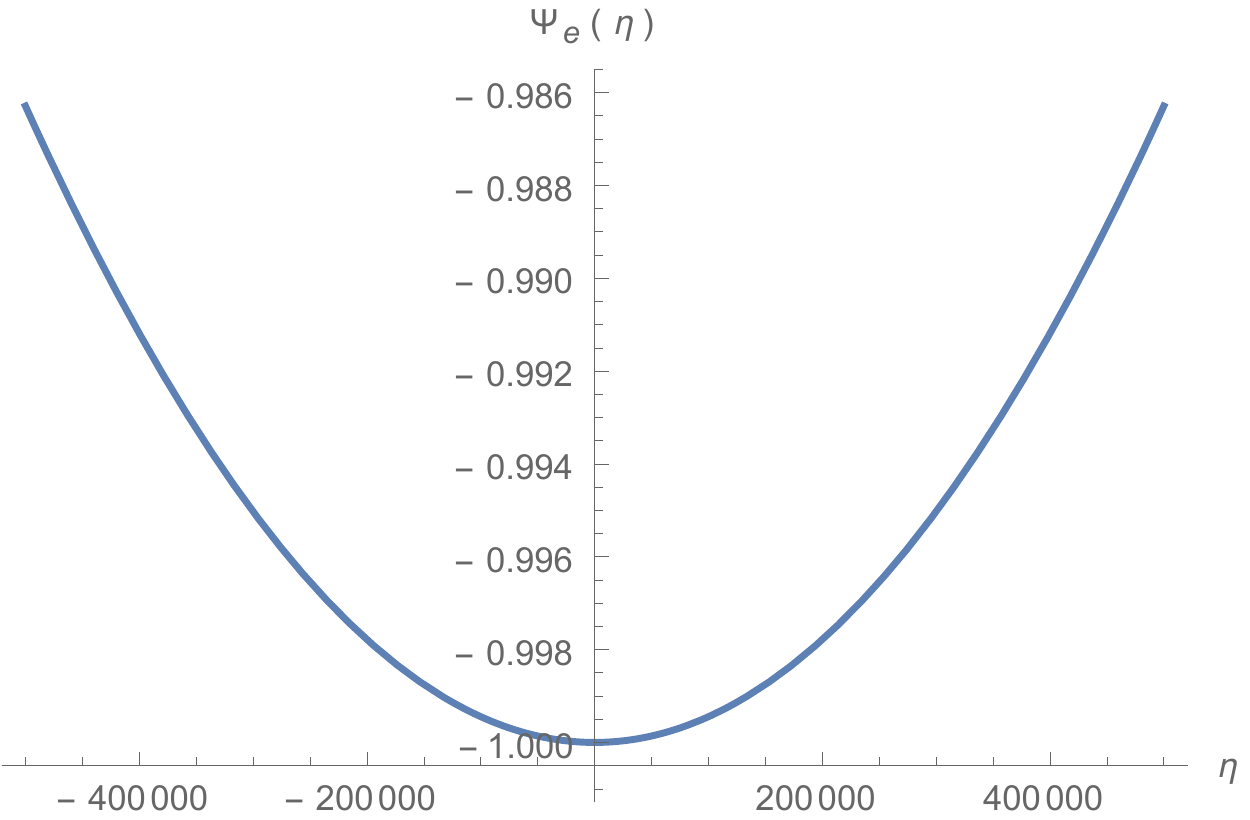}
\caption{Time evolution of $\Psi_e$ in the Jordan frame for the
  initial conditions $\Phi_e(0)=-10^{-2}$, $\Psi_e(0)=-1$,
  $\Phi_e'(0)=0$, $\Psi_e'(0)=0$. The background evolution is the one
  obtained by assuming the same initial conditions as mentioned in the
  caption of Fig.\ref{arad1}.}
\label{Psi4}
\end{minipage}
\end{figure}

Let us discuss these even solutions in more detail. Suppose
these perturbations start with an absolute value less than unity prior
to the bounce. If we want to guarantee that the perturbations indeed
always remain in the perturbative level then one can specify a
general condition as follows.
\begin{figure}[t!]
\begin{minipage}[b]{0.5\linewidth}
\centering
\includegraphics[scale=.5]{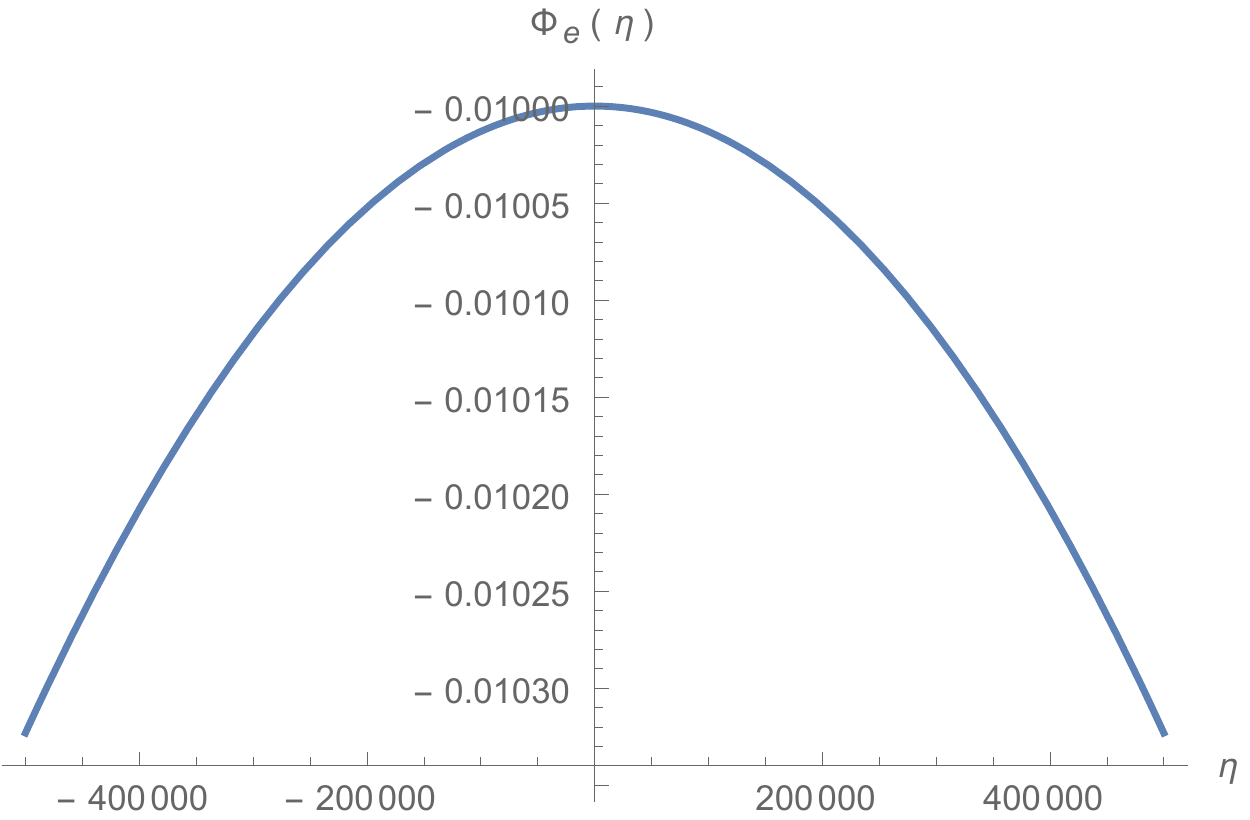}
\caption{Time evolution of $\Phi_e$ in the Jordan frame for the
  initial conditions $\Phi_e(0)=-10^{-2}$, $\Psi_e(0)=10^{-3}$,
  $\Phi_e'(0)=0$, $\Psi_e'(0)=0$. The background evolution is the one
  obtained by assuming the same initial conditions as mentioned in the
  caption of Fig.\ref{arad1}.}
\label{Phi2}
\end{minipage}
\hspace{0.2cm}
\begin{minipage}[b]{0.5\linewidth}
\centering
\includegraphics[scale=.5]{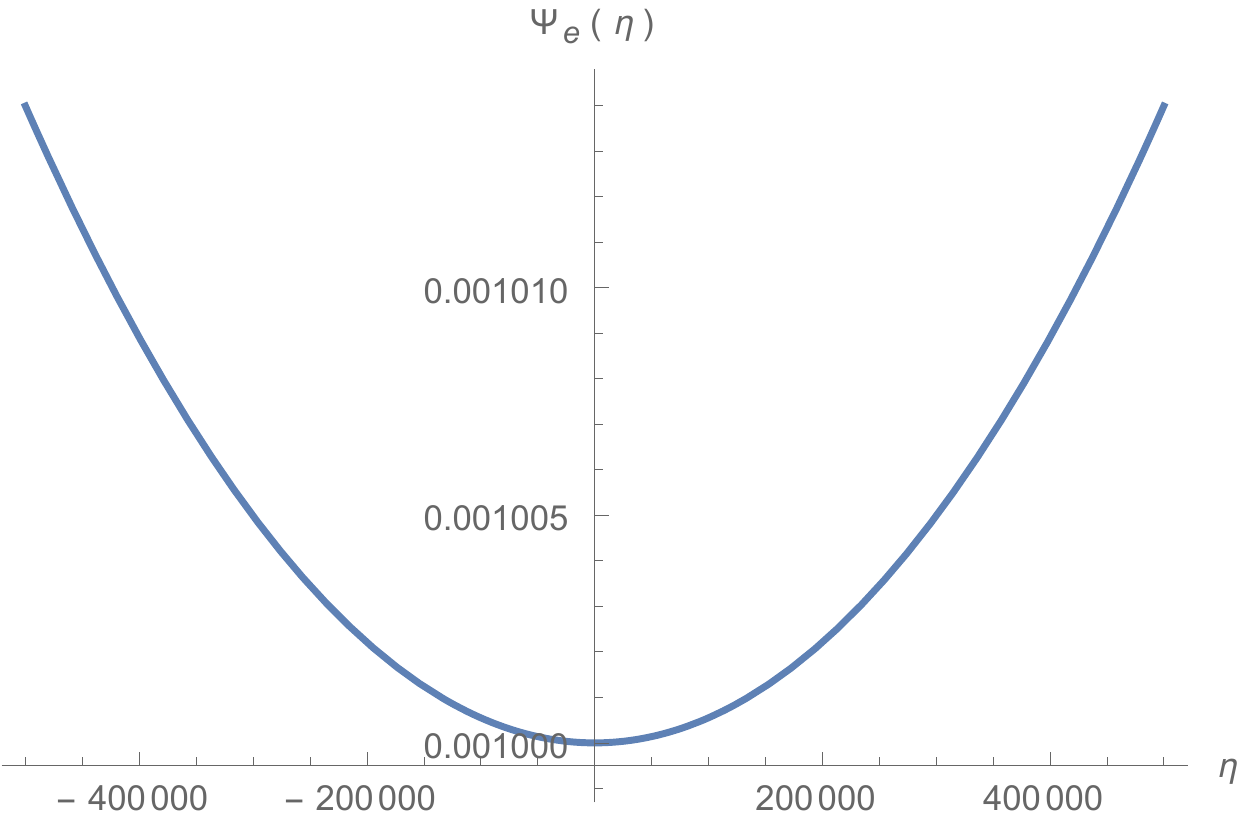}
\caption{Time evolution of $\Psi_e$ in the Jordan frame for the
  initial conditions $\Phi_e(0)=-10^{-2}$, $\Psi_e(0)=10^{-3}$,
  $\Phi_e'(0)=0$, $\Psi_e'(0)=0$. The background evolution is the one
  obtained by assuming the same initial conditions as mentioned in the
  caption of Fig.\ref{arad1}.}
\label{Psi2}
\end{minipage}
\end{figure}
If the relevant perturbation ($\Phi_e(\eta)$ or $\Psi_e(\eta)$) is in
the perturbative region (meaning $|\Phi_e(\eta)|<1$ or
$|\Psi_e(\eta)|<1$) during the contracting phase then the perturbation
will remain perturbative throughout the bounce if the minimum of the
function has a value greater than $-1$ at $\eta=0$ or the maximum of
the function is less than $+1$ at $\eta=0$.  A major part of the above
condition is encoded in the following inequalities
\begin{enumerate}
\item $\Phi_e(0)\geq 0$, $\Phi_e''(0)>0$, $\Psi_e(0)\geq 0$, $\Psi_e''(0)>0$.
\item $\Phi_e(0)\geq 0$, $\Phi_e''(0)>0$, $\Psi_e(0)\leq 0$, $\Psi_e''(0)<0$.
\item $\Phi_e(0)\leq 0$, $\Phi_e''(0)<0$, $\Psi_e(0)\leq 0$, $\Psi_e''(0)<0$.
\item $\Phi_e(0)\leq 0$, $\Phi_e''(0)<0$, $\Psi_e(0)\geq 0$, $\Psi_e''(0)>0$.
\end{enumerate}
The above inequalities only form a subset of the various cases which
may arise from the condition stated earlier.  The above inequalities
can be used for a simplified consideration of the problem at hand.
Solutions for perturbations are found out by solving two coupled
differential equations, each having the form as in
Eq.~(\ref{ptbd_generic}). Since $\Phi_e'(0)=\Psi_e'(0)=0$, the two
perturbation evolution equations written at $\eta=0$ relates the four
quantities $\Phi_e(0)$, $\Psi_e(0)$, $\Phi_e''(0)$, $\Psi_e''(0)$ by
two simultaneous algebraic equations. For the numerical initial
conditions that we have chosen (as mentioned in the caption of
Fig.~\ref{arad1}) and for $k=10^{-8}$ these two simultaneous algebraic
equations are as follows:
\begin{eqnarray}
1.82\Phi''(0)-4.11\times 10^{-13}\Phi(0)+1.82\Psi''(0)+4.60\times 10^{-13}\Psi(0)&=&0.\\
-0.20\Phi(0)-9.03\times 10^{12}\Psi''(0)-\Psi (0)&=&0.
\end{eqnarray}
These two algebraic equations can then be solved to obtain $\Phi_e''(0)$, $\Psi_e''(0)$ in
terms of $\Phi_e(0)$, $\Psi_e(0)$ as follows
\begin{eqnarray}
\Phi_e''(0)=2.48\times 10^{-13}\Phi_e(0)-1.42\times 10^{-13}\Psi_e(0),\\
\Psi_e''(0)=-0.22\times 10^{-13}\Phi_e(0)-1.11\times 10^{-13}\Psi_e(0).
\end{eqnarray}
Let us now check for the four conditions mentioned above which
guarantee that if the perturbations started with a perturbative value,
then they remain perturbative throughout the bounce. Straightforward
check gives that the first condition and the third condition can never
be satisfied, at least in the simplified cases we are discussing. The
second condition and the fourth condition can be satisfied
respectively for
\begin{center}
  $\Phi_e(0)>0\,\,,\,\,-0.2\Phi_e(0)<\Psi_e(0)\leq 0$ and $\Phi_e(0)<0\,\,,\,\,
  0\leq\Psi_e(0)<-0.2\Phi_e(0)$.
\end{center}
Figs.~(\ref{Phi1},\ref{Psi1}) correspond to the first of the above
conditions and Figs.~(\ref{Phi2},\ref{Psi2}) correspond to the
second. As is clearly seen from the plots, for both of these
conditions produce perturbative evolution of the even Bardeen
potentials through the bounce. The various other cases of perturbation evolution are shown in
Figs.\ref{Phi3},\ref{Psi3},\ref{Phi4},\ref{Psi4}.

A general solution of the perturbation evolution equations
$c_1\Phi_e+c_2\Phi_o$ and $c_1\Psi_e+c_2\Psi_o$ in general will not
vanish or have a local extremum at the bounce point. But since
$|\Phi_o|$, $|\Psi_o|$ must vanish and $|\Phi_e|$, $|\Psi_e|$ must
have local extrema at the bounce point, we can opine about the nature
of the perturbations at the bounce point by noting the behavior of
the even potentials only. For the perturbation theory to be valid
throughout the bouncing time period, the perturbations need to remain
always at a perturbative level. The work out in this brief section
shows that there are various cases where the perturbations in the
longitudinal gauge remains perturbative throughout the bounce. We do
not rule out cases where the perturbations may become non-perturbative
near the bounce but our analysis shows that the perturbations do not
diverge near the bounce point. 

Before we end this section we want to present a discussion on the
instabilities present in general bouncing models based on GR with
modified matter sector (as ghosts, Galileon) and $f(R)$ theories with
standard matter sector. In GR based theories the non standard matter
component violates the null energy condition (NEC) near the bounce
point and consequently this kind of matter component is essential for
bounce in spatially flat FLRW models. But this NEC breaking phase may
produce instabilities as ghost instabilities. One may eradicate the
ghost instability by considering bounce in Galileon theory or in
general Horndeski theory. The Horndeski like theories show gradient
instabilities due to which the sound speed squared turns out to be
negative and there is an exponential growth of perturbations. The
following Refs.\cite{Cai:2012va, Kobayashi:2016xpl, Cai:2016thi} give
a detailed discussion on the issues of instabilities arising in GR
based bounces and methods to eradicate them. In $f(R)$ bounce theories
one generally do not employ non-standard matter, the energy density
and pressure for the non-standard matter component is produced from
the scalar curvature and encoded in $\rho_{\rm eff}$ and $P_{\rm
  eff}$, as shown in Eqs.~(\ref{fried}) and (\ref{2ndeqn}). The matter
part specified by $\rho$ and $P$ generally always has $\rho+ P >
0$. During bounce the curvature contribution (to energy density and
pressure) acts like non-standard matter in $f(R)$ theory.  In
Ref.\cite{Sotiriou:2008rp} the authors show that in $f(R)$ theory,
which is a higher derivative theory, ghosts do not appear.  Although
in exponential gravity one naturally takes care of the conditions
$f'(R)>0$ and $f''(R)>0$ which eradicates inherent instabilities of
$f(R)$ dynamics there may remain some hidden form of instabilities
corresponding to the gradient instability which may make some of the
scaler modes to become non-perturbative during the time of bounce. The
bounce problem in $f(R)$ theories are still in their infancy and we
hope in the near future a stability analysis of the perturbation modes
will be presented in full detail.
\subsection{Brief comments about tensor perturbations through bounce}

The calculations on tensor perturbations in $f(R)$ gravity, in the context of inflation, has been done in Ref.\cite{DeFelice:2010aj}. From the above reference one can see that in general the tensor perturbation evolution equation, in coordinate time, is given by 
\begin{equation}
\ddot{h}_\lambda +\frac{\dot({a^3 F})}{(a^3 F)} \dot{h}_\lambda+ \frac{k^2}{a^2} h_\lambda =0
\end{equation}
for $\lambda=+,\times$, which are the polarization modes of the tensor
perturbations. Here $F=df/dR$. The evolution equation for the tensor
perturbations follows the same generic form of perturbation equation
presented in Eq.~(\ref{ptbd_generic}), as the coefficient function
multiplying $\dot{h}_\lambda$ is odd and the coefficient function
multiplying $h_\lambda$ is even in coordinate time. This shows that in
principle one can produce a similar analysis, of the tensor
perturbations in the present case, based on the symmetry of the
gravitational wave amplitudes in time. We do not expect any
singularity of the tensor modes as the above equation does not have
any singular point in the bouncing regime.  In
Ref.~\cite{Raveendran:2017vfx} the authors showed that the two scalar
field matter bounce model in GR produces a tensor-to-scalar ratio($r$)
that is within the observed upper bound. In f(R) bounce models,
however, explicit analytical calculation of tensor-to-scalar ratio is
still missing. We expect that the effect of modified $f(R)$ gravity
models are valid near the vicinity of the bounce and if the previous
history of the universe does not produce large gravitational power
spectrum (compared to comoving curvature power spectrum) then
exponential gravity will not amplify $r$ (the tensor-to-scalar
ratio). In Starobinsky inflationary scenario where an $R^2$ correction
term is incorporated in the Lagrangian, $r$ comes out to be smaller
than the observational bound\cite{DeFelice:2010aj}. Also, in
Ref.\cite{Odintsov:2017qif} the authors consider inflation in
exponential gravity and calculate $r$ to be below the observational
limit.  From these facts, we can expect that inclusion of such
correction terms in the Lagrangian will not abruptly increase the
value of $r$ (from observational bound). This point needs explicit
analytical consideration, which we hope to address in future
publications.

In the next section we will discuss some interesting exact solutions
in exponential gravity. The solutions plotted in this section are
obtained numerically, rarely in $f(R)$ gravity theory we have the
privilege of having exact solutions\cite{Schmidt:1998sn, Wei:2015xax}.
The solutions given in the next section do not define the form of
$f(R)$, the solutions are exact solutions of exponential gravity.
\section{Two exact solutions in exponential gravity}
\label{tes}

In this section we show that $(1/\alpha)e^{\alpha R}$ gravity has both
an exact bouncing solution and a solution with constant Ricci
curvature. As we are not using the concept of conformal time
explicitly in this section we go back to our old convention where
derivatives with $R$ are specified by primes.  We will show that the
constant curvature(de Sitter point) solution admits an inflationary
scale-factor. We have not worked out the cosmology around the de
Sitter point and the analogy with the inflationary scale-factor may
turn out to be purely formal.  Since out of the three equations
Eq.~(\ref{fried}), Eq.~(\ref{2ndeqn}) and Eq.~(\ref{cont}), only two
are independent, we can choose to work with equations
Eq.~(\ref{fried}) and Eq.~(\ref{cont}) for the sake of
convenience. Using the standard solution $\rho=\rho_0
a^{-3(1+\omega)}$ obtained from equation Eq.~(\ref{cont}) where
$\rho_0$ and $\omega$ are constants, the constraint equation
Eq.~(\ref{fried}), can be written as
\begin{equation}
6H^2f^{\prime}=2\kappa\rho_0e^{-3(1+\omega)\ln a}+Rf^{\prime}-f-6H\dot{R}f^{\prime\prime}.
\end{equation}
For $(1/\alpha)e^{\alpha R}$ gravity, this becomes
\begin{equation}
6H^2-R+6\alpha H\dot{R}+\frac{1}{\alpha}=2\kappa\rho_0e^{-3(1+\omega)\ln a +\alpha R}.
\label{fried_exp}
\end{equation}
Using the above equation let us now prove the existence of exact
exponential bouncing solution and exact de-Sitter solution.
\subsection{Exact exponential bouncing solution}

In this subsection we choose a scale-factor $a(t)=e^{Ct^2}$, where $C$
is a positive constant. In such a case we get
\begin{center}
$H(t)=2Ct\,\,,\,\,R=12C(1+4Ct^2)\,\,,\,\,\dot{R}=96C^2t$.
\end{center}
Using all these information and the modified constraint equation,
Eq.~(\ref{fried_exp}), we get
\begin{equation}
\left(\frac{1}{\alpha}-12C\right)+24C^2t^2\left(48\alpha C-1\right)=2\kappa\rho_0
\exp\left[-12\alpha C -3Ct^2(1+\omega+16\alpha C)\right].
\end{equation}
If $a(t)=e^{Ct^2}$ is an exact solution of $(1/\alpha)e^{\alpha R}$
gravity theory, then the above equation has to be satisfied for all
values of $t$. This condition can be satisfied when
\begin{center}
$\alpha C=\frac{1}{48}\,\,,\,\,\omega=-\frac{4}{3}$\,.
\end{center}
We see that for these values of the parameters $\alpha$, $C$ and
$\omega$, $\rho_0$ is given by
\begin{center}
$\rho_0=\frac{3e^{1/4}}{8\kappa\alpha}$,
\end{center}
which is always positive. Exponential gravity can yield a bouncing
universe with exponential scale-factor in presence of hydrodynamic
matter which can have negative pressure but whose energy density is
positive definite. The causal nature of the universe where the
scale-factor is as given in this section is worked out in \cite{Bhattacharya:2017evz}.
\subsection{Exact de-Sitter solution}

A de-Sitter solution is a vacuum solution of constant positive
curvature in GR. We use the same terminology and name our solution as
de Sitter solution, as our solution in $f(R)$ gravity is a constant
positive curvature solution in presence of positive vacuum energy. We
assume that the scale-factor of the universe as $a(t)=e^{Ht}$, where
$H$ is a positive constant and consequently
\begin{center}
$R=12H^2$.
\end{center}
Using the above information in the modified constraint equation,
Eq.~(\ref{fried_exp}), and setting $\rho_0=0$, we get
\begin{equation}
H^2=\frac{1}{6\alpha}\,\,\,\,\rm{or}\,\,\,\,R=\frac{2}{\alpha}.
\end{equation}
Therefore $(1/\alpha)e^{\alpha R}$ gravity has an exact de Sitter
solution in which the constant value of the Ricci scalar is given by
$R=2/\alpha$. In the Einstein frame this value of $R$ correspond to
$\phi=2\sqrt{3/(2\kappa)}$, the point at which the potential $V(\phi)$
assumes it's maximum.
\begin{figure}
\centering
\includegraphics[scale=.7]{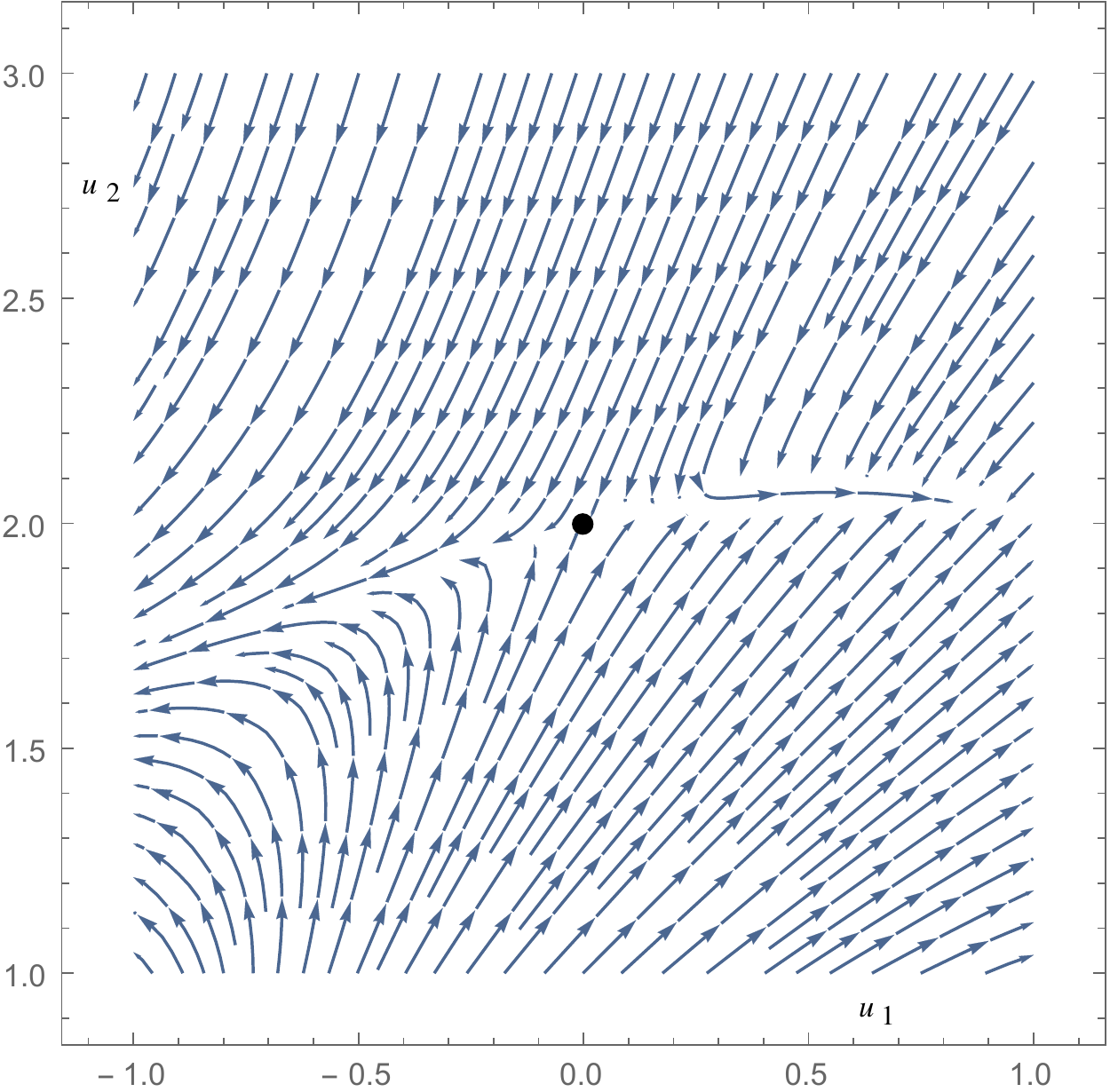}
\caption{Phase space plot in the $u_1u_2$ plane of isotropic vacuum
  solutions in exponential gravity, where $u_1$ is along the $x-$axis
  and $u_2$ is along the $y-$axis. The de-Sitter solution is given by
  the point $(0,2)$, which is represented by the dot at the center of the figure. The arrows
  show the direction of the solution flow. The phase flows clearly
  show that the de-Sitter solution is a saddle fixed point.}
\label{phase_plot}
\end{figure}

As the de-Sitter point lies at the top of the Einstein frame potential
one can intuitively conclude that the de Sitter solution is an
unstable solution. In the rest of the section we will show that the de
Sitter solution is indeed an unstable solution.  To analyze the
stability of this solution, we resort to a dynamical system analysis
in the Jordan frame in terms of normalized, dimensionless dynamical variables as
formulated in references \cite{Amendola:2006we, Carloni:2007br,
  Carloni:2004kp, Odintsov:2017tbc}. Let us define the dimensionless
dynamical variables
\begin{equation}
u_1=\frac{\alpha\dot{R}}{H}\,\,,\,\,u_2=\frac{R}{6H^2}\,\,,\,\,u_3=\frac{1}{6\alpha H^2}.
\end{equation}
Eq.~(\ref{fried_exp}) then implies the following constraint between them
\begin{equation}
u_1-u_2+u_3=-1,
\end{equation}
which implies only two of them are independent. Note that for the de-Sitter solution
\begin{equation}
u_1=0\,\,,\,\,u_2=2\,\,,\,\,u_3=\frac{1}{6\alpha H^2}.
\end{equation}
Therefore for the constraint equation to be satisfied one must have
$H^2=1/(6\alpha)$, which is consistent with what we previously
obtained. Without loss of generality, we can take $u_1$ and $u_2$ as
the two independent dynamical variables. With the help of the
Eq.~(\ref{2ndeqn}), we can find the dynamical equations for $u_1$,
$u_2$ in terms of the dimensionless logarithmic time variable
$N\equiv\ln a$ as:
\begin{eqnarray}
\frac{du_1}{dN}&=&4+3u_1-2u_2-u_1^2-u_1u_2,\\
\frac{du_2}{dN}&=&-u_1-u_1^2+4u_2+u_1u_2-2u_2^2.
\end{eqnarray}
It is straightforward to check that the de-Sitter point given by the
coordinates $(0,2)$ in the $u_1u_2$ plane is a fixed point, i.e., at this point
\begin{center}
$\frac{du_1}{dN}=\frac{du_2}{dN}=0$.
\end{center} 
The eigenvalues of the Jacobian matrix at this point are
\begin{center}
$\frac{1}{2}(-3\pm\sqrt{17})$.
\end{center}
The first eigenvalue is positive and the second is negative, meaning
the de-Sitter point is a saddle point in the space of isotropic vacuum
solutions in exponential gravity. Fig.~(\ref{phase_plot}) shows the
flows of the solutions in the phase space. 

After discussing the exact solutions in exponential gravity we discuss
the extra solutions we get in exponential gravity if we allow
exponential gravity to cross $R=0$ value. In general we expect the
theory to be similar to GR near $R=0$ and the $f(R)$ effect becomes
effective for high values of $R$. But if we assume that $f(R)$ gravity
can be also used for small $R$ values then we come across new kind of
solutions presented in the next section.
\section{New solutions in exponential gravity}
\label{new}

\begin{figure}[t!]
\begin{minipage}[b]{0.5\linewidth}
\centering
\includegraphics[scale=.6]{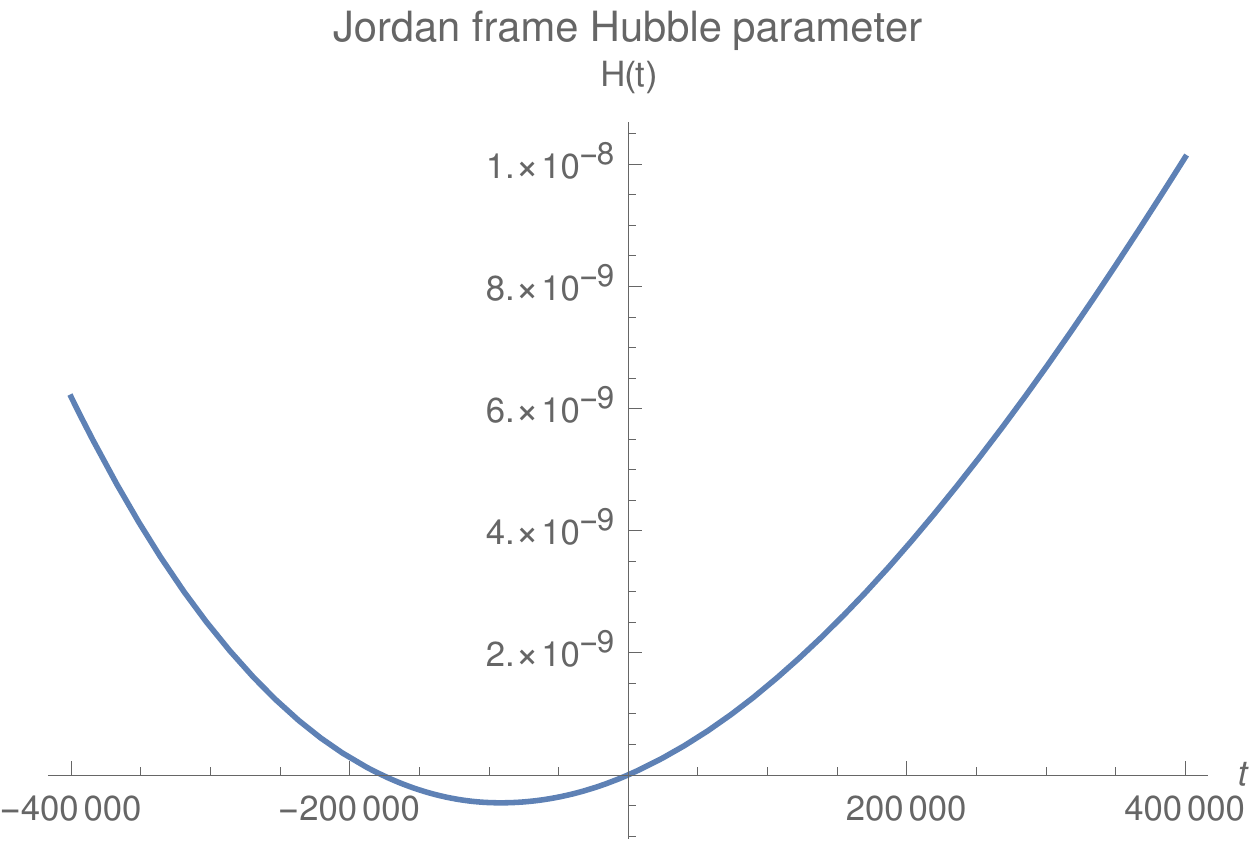}
\caption{Time evolution of the Hubble parameter in the Jordan frame
 in radiation background in exponential gravity with the conditions 
$H(0)=0$, $\dot{H}(0)=10^{-14}$, $\ddot{H}(0)=10^{-19}.$} 
\label{hrad3}
\end{minipage}
\hspace{0.2cm}
\begin{minipage}[b]{0.5\linewidth}
\centering
\includegraphics[scale=.9]{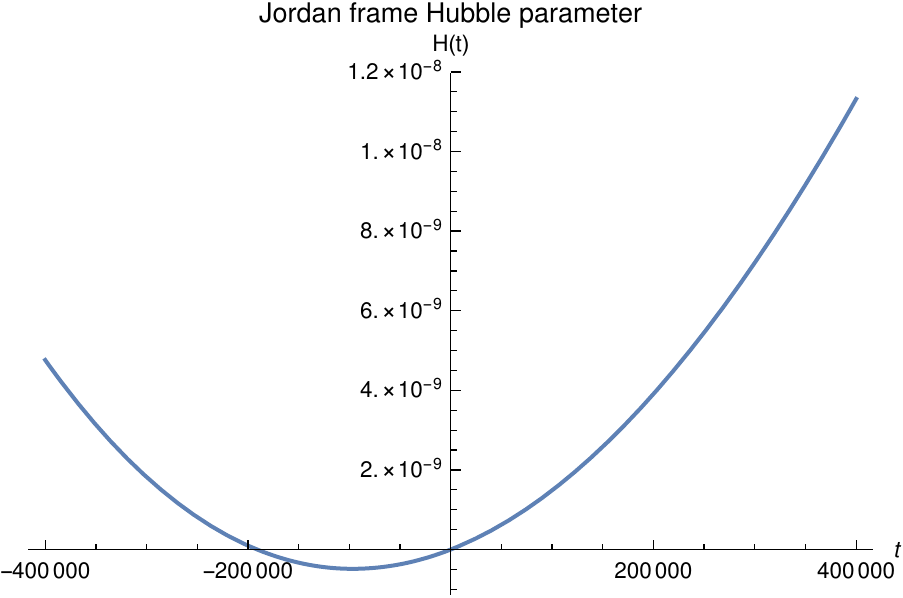}
\caption{Time evolution of the Hubble parameter in the Jordan frame
in vacuum background in exponential gravity with the initial conditions 
$H(0)=0$, $\dot{H}(0)=10^{-14}$, $\ddot{H}(0)=10^{-19}.$}
\label{hvac3}
\end{minipage}
\end{figure} 

If we allow the cosmological model presented in this article to be
valid for very small $R$ then we get new class of solutions which, to
our understanding, was never reported before in any discussions about
$f(R)$ gravity. The important feature of the solutions presented in
this section is the generality of the discussion. Most of the results
we present in this section is generally true for any $f(R)$ theory
which accommodates a cosmological bounce. We present the results for
exponential gravity in the present paper.

\begin{figure}[t!]
\begin{minipage}[b]{0.5\linewidth}
\centering
\includegraphics[scale=.6]{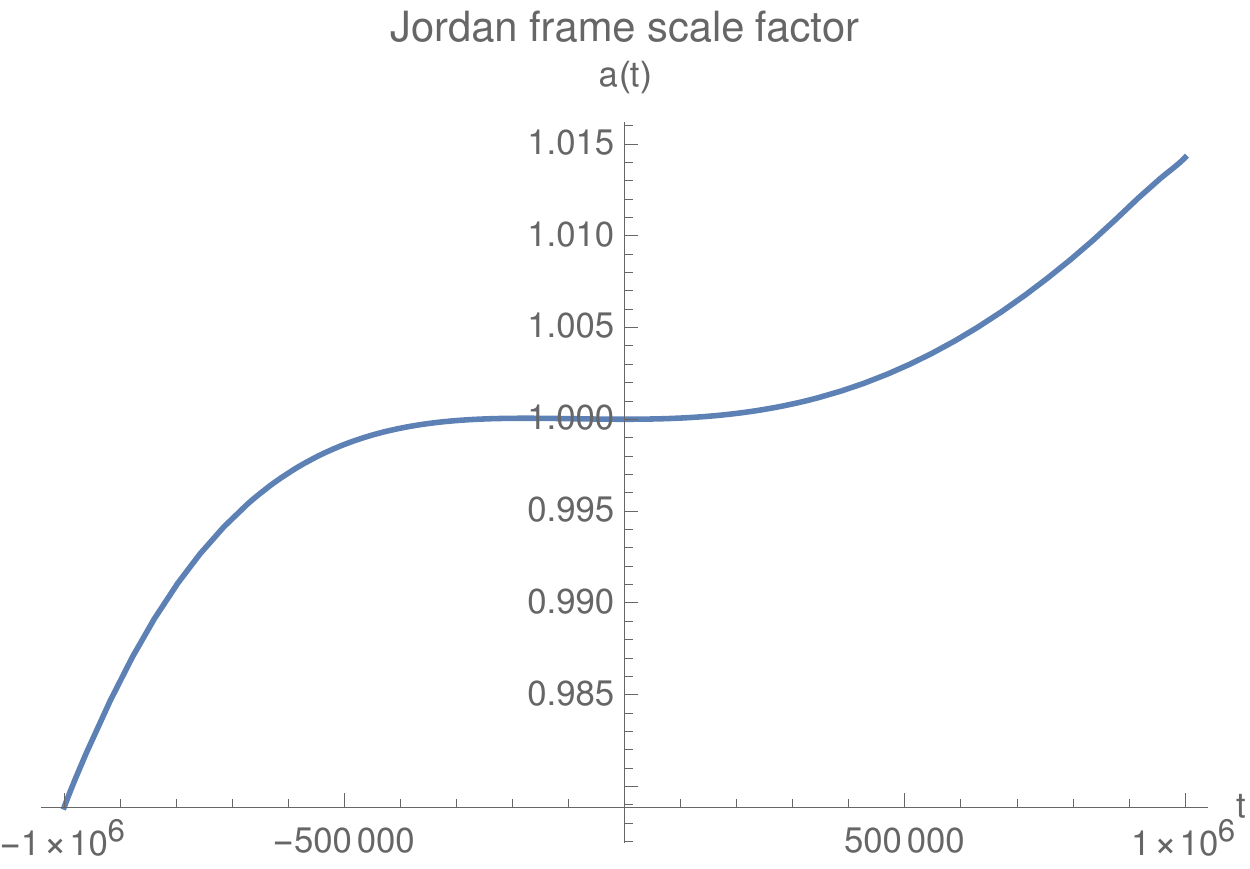}
\caption{Time evolution of the scale factor in the Jordan frame in radiation background 
in exponential gravity found with the  conditions
$H(0)=0$, $\dot{H}(0)=10^{-14}$, $\ddot{H}(0)=10^{-19}.$} 
\label{narad3}
\end{minipage}
\hspace{0.2cm}
\begin{minipage}[b]{0.5\linewidth}
\centering
\includegraphics[scale=.7]{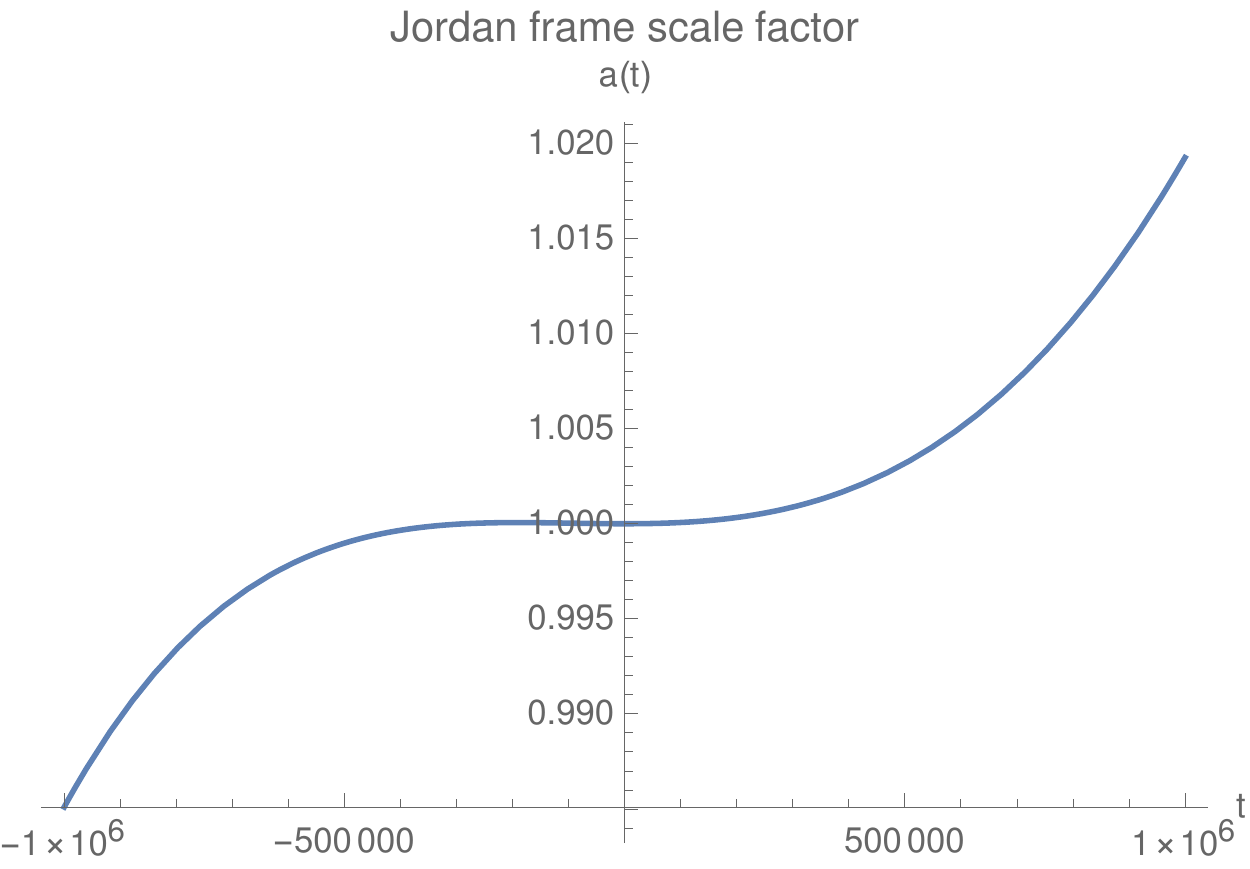}
\caption{Time evolution of the scale factor in the Jordan frame in
  vacuum background in exponential gravity found with the conditions
  $H(0)=0$, $\dot{H}(0)=10^{-14}$, $\ddot{H}(0)=10^{-19}.$}
\label{navac3}
\end{minipage}
\end{figure}

For a cosmological bounce one requires $H(0)=0$ and $\dot{H}(0)>0$ in
the Jordan frame. In $f(R)$ gravity we can give another intermediate
condition, and this condition is on $\ddot{H}(0)$. Suppose we specify 
$\ddot{H}(0)=\epsilon$ where $\epsilon$ is a real
constant. Near the point where $t=0$ we can then write approximately
$\dot{H}(t) \approx \epsilon t + b$ where $b$ is a positive, real constant. 
The approximation remains valid as long as $\dot{H}$ varies linearly
near the bouncing point.  In such a case we can integrate once more
and write the expression of the Hubble parameter near the bounce point
as
\begin{eqnarray}
H(t) \approx \frac12 \epsilon t^2 + b t\,,
\label{htj}
\end{eqnarray}
the integration constant has been set to zero because the Hubble
parameter vanishes at $t=0$. The above equation shows that $H$ can be
zero at two separate time instants, given by 
\begin{eqnarray} 
t=0\,,\,\,\,\,\,t \approx -\frac{2b}{\epsilon}\,,
\label{tvals}
\end{eqnarray}
The above statements are in general true for any $f(R)$ theory which
accommodates a cosmological bounce. The system behavior, in the
particular case of exponential gravity, is shown in the
Fig.~\ref{hrad3}, which depicts an universe filled up radiation, and
in Fig.~\ref{hvac3}, which depicts the properties of matter less
universe. In both the cases we see that the Hubble parameter reaches
zero value at two time instants, as predicted from
Eq.~(\ref{tvals}). One of the points when the Hubble parameter is zero
is placed at $t=0$ where as the other time instant is approximately
$-2 \times 10^5$, of the same order as predicted from
Eq.~(\ref{tvals}). If one wants to see how the scale-factor of the
universe behaves during this period then one can look at Fig.~\ref{narad3} and
Fig.~\ref{navac3}. From both the figures one can see that initially the universe was
expanding and this expansion slowly stops and a brief period of
contraction sets in, the period when the Hubble parameter turns
negative, and then this contraction stops slowly and the universe
again enters a period of expansion. It is to be noted that if we
increase the value of $\epsilon$, or $\ddot{H}(0)$ in Jordan frame, then
the temporal separation of the two roots in Eq.~(\ref{tvals})
decreases. This implies that the initial expansion phase ends near to
the point where later expansion phase starts. In the asymmetric bouncing
solutions, as depicted in Fig.~\ref{arad3} and Fig.~\ref{avac3}, in
section \ref{begs} we have used non-zero values of $\ddot{H}(0)$ in
the Jordan frame, still we are getting perfect bounces and not
oscillatory behavior as expected from our present analysis. The reason
for this is related to the time period used to study the asymmetric
bounces in section \ref{begs} and the specific value of $\ddot{H}(0)$
used there. From the conditions given for the asymmetric bounces one
can easily show that the oscillatory behavior should have been evident
if we presented the plot for a larger time period. In the time period
of interest, only the bouncing region becomes prominent in
Fig.~\ref{arad3} and Fig.~\ref{avac3}.

Before we conclude this section we want to show another important
property of the new solutions shown. For flat FLRW solutions one can
write the Ricci scalar as
\begin{eqnarray} 
R = 6 (\dot{H} +2 H^2) = 6(\epsilon t + b)+12\left(\frac{\epsilon t^2}{2} +bt\right)^2\,, 
\end{eqnarray}
close to the bounce point. From the above expression one can easily
verify that
$$R(0)=6b\,,\,\,\,\,\,\,R(t=-2b/\epsilon)=-6b\,,$$ which shows that
the Ricci scalar has to change sign in between the two temporal values
when the Hubble parameter vanishes. As a result of this fact, the
universe has to cross the point when $R=0$, between the two expanding
regimes.  This result is a general result and is true for any $f(R)$
theory which accommodates a cosmological bounce. If we want to work
with large, positive values of $R$ and treat $f(R)$ theory as an
effective gravitational theory then the new solutions become redundant
as only a bounce exists, as analyzed in section \ref{begs}. On the
other hand if one wants to keep the effect of $f(R)$ bouncing
cosmology for very small values of the Ricci scalar, then the new
solutions will show up.
\section{Conclusion}

In this article we presented some early universe solutions coming out
from exponential $f(R)$ gravity. The primary reason for choosing
exponential gravity is related to the stability of the theory. For
positive cosmological constant one always gets $f^\prime(R)>0$ and
$f^{\prime \prime}(R)>0$ for all values of $R$. We presume the theory
to be unstable when one moves away from bounce point as in the
relevant time limit the Ricci scalar becomes negative and remains
unbounded from below. This feature of the theory is well represented
in the Einstein frame potential, as shown in Fig.~\ref{pot_phi}. We
will like to interpret the early universe results, coming out from
exponential $f(R)$ theory, as an outcome from some effective theory of
gravity which affects in the ultraviolet end.

In this paper we initially present the basics of cosmological dynamics
in the two conformal frames, the Jordan frame where the original
problem is posed and the Einstein frame which is used to calculate the
dynamical development of the system. Our results related to bouncing
cosmologies uses two conformal transformations. We show that it is
difficult in $f(R)$ theory to have stable bounces. In this regard
exponential gravity is an interesting exception as it satisfies the
stability conditions for all values as $R$. In section \ref{dbscen} we
present the bouncing condition in the Jordan frame and its
corresponding condition in the Einstein frame. We want to specify here
that in spatially flat FLRW spacetimes one in general does not get a
simultaneous bounce in both the conformal frames when matter in Jordan
frame satisfies $\rho + P \ge 0$ \cite{Paul:2014cxa}. The conditions
of bounce in Jordan frame, when written down in the Einstein frame,
shows that the conditions depend upon the second time derivative of
the Jordan frame Hubble parameter. Consequently, the bouncing
intermediate conditions actually involves the values of all the
relevant time derivatives of $H$ in the Jordan frame. It is to be
noted that unlike $f(R)$ theories, GR based cosmology does not require
the specification of $\ddot{H}$ at any instant of time. In this
article we point out specifically how $\ddot{H}$ can affect
cosmological dynamics in higher derivative gravity theory. The
evolution of scalar metric perturbations through bounce is presented
in section \ref{ptb}. The whole analysis is done in the Jordan frame
and it is shown that the perturbations do not attain any singularity
at the bounce point although there me be some modes which can become
non-perturbative very near the bounce point. The reason for such
non-perturbative evolution may be related to the behavior of curvature
related terms near the bounce point. A thorough analysis of stability
of metric perturbations in $f(R)$ cosmology, near bounce, is not yet
present we hope the theory will be formulated in the near future. We
also briefly opine on the fate of the tensor-to-scalar ratio in our
$f(R)$ theory induced bounce. We do not expect the present theory to
amplify the ratio compared to its value in the prebounce phase.

In the article we present the bouncing solutions in the Jordan
frame. In the bouncing solution calculations we use the Einstein frame
as an auxiliary frame where the main calculation is done and then we
transport the solutions in the Jordan frame via a conformal
transformation. The bouncing solutions presented in this paper involve
both bounce in vacuum and bounce in the presence of hydrodynamic
matter. The examples involve both symmetric and asymmetric
bounces. The asymmetry of the bounces is related to finite values of
$\ddot{H}(0)$ in the Jordan frame. For symmetric bounces
$\ddot{H}(0)=0$. The solutions presented in section \ref{begs} are
computed numerically and they are well behaved in our time period of
interest, $-10^5 \le t \le 10^5$. Outside the time window the
solutions can show other features. All the bounces concerned in
section \ref{begs} involves variation of positive values of the Ricci
scalar in the Jordan frame. Treated $f(R)$ theory as an effective
theory of gravity, the exponential gravity bounces become more
plausible at higher positive values of $R$. As exponential gravity
becomes similar to GR with positive cosmological constant for low $R$
values one may like to infer that GR effects becomes stronger as $R$
nears zero. The exact turn over from $f(R)$ to GR may involve new
physics and is beyond the scope of this paper.

Another interesting property of exponential gravity, as discussed in
the paper, is the existence of exact solutions. We present two such
solutions in the article. One exact solution is a bouncing solution
where the scale-factor of the universe is given by an exponential
function of the square of Jordan frame time. This solution can be
realized in an universe with matter having positive energy density and
equation of state as $-4/3$.  With an equation of state lesser than
$-1$, the matter does not satisfy $\rho + P \ge 0$ in the Jordan frame
and consequently for such an universe one may expect simultaneous
bounces in both the Jordan frame and the Einstein frame. The other
exact solution is the exponential expansion solution with constant $H$
at a de Sitter point. In vacuum, exponential gravity allows such a
solution to exist. One can easily show that such a de-Sitter point
exists because exponential gravity involves a positive cosmological
constant. If the form of $f(R)$ is suitably changed such that it does
not contain any cosmological constant the de-Sitter point
vanishes. Using the techniques of dynamical systems we have shown
that a constant Hubble parameter solution at the de-Sitter point in
exponential gravity is an unstable solution. In effect it is a saddle
point solution in the phase space of suitable defined dimensionless
phase space variables. We do not present the dynamical system approach
in the other solutions in this paper because all the other solutions
involve the values $H=0$ and the phase space variables in our analysis
always have the Hubble parameter in the denominator. We hope to
construct a suitable dynamical systems approach to tackle bouncing problems
in the near future. 

We present a new solution in $f(R)$ gravity theory in the penultimate
section. The new solution in $f(R)$ theories are allowed only if the
theory is allowed to be unmodified in the low $R$ regime. As our
theory is stable it can safely be extended to the low $R$ regime. The
only cost one has to pay to attain these new solutions is that one has
to reject the point of view that $f(R)$ gravity is an ultraviolet
modification of GR effects. The new results are related to non-zero
values of $\ddot{H}(0)$ in the Jordan frame when the other bouncing
conditions hold. In such a case one can a have a solution which
represent decelerated expansion of the universe in the past. At some
point in the past the decelerated expansion comes to a halt momentarily
and contraction of the universe starts. This contraction does not lead
to a spacetime singularity. In time this contraction slows down and
the universe comes to a static configuration momentarily after which
again the universe starts to expand. This solution is practically not
a bouncing solution, although one can get this solution with an extra
intermediate condition on top of the bouncing conditions at $t=0$. The
new result which we obtain in this paper is a general result in $f(R)$
gravity which accommodates a bounce. We explicitly show the nature of
the solutions in exponential gravity.


\begin{thebibliography}{10}
\bibitem{Kragh:2013dva}
  H.~Kragh,
  arXiv:1308.0932 [physics.hist-ph].

\bibitem{Guth:1980zm} 
  A.~H.~Guth,
  Phys.\ Rev.\ D {\bf 23}, 347 (1981).
  doi:10.1103/PhysRevD.23.347

\bibitem{Linde:1981mu} 
  A.~D.~Linde,
  Phys.\ Lett.\  {\bf 108B}, 389 (1982).
  doi:10.1016/0370-2693(82)91219-9

\bibitem{Martin:2013tda} 
  J.~Martin, C.~Ringeval and V.~Vennin,
  Phys.\ Dark Univ.\  {\bf 5-6}, 75 (2014)
  doi:10.1016/j.dark.2014.01.003
  [arXiv:1303.3787 [astro-ph.CO]].

\bibitem{Borde:1996pt}
  A.~Borde and A.~Vilenkin,
  Int.\ J.\ Mod.\ Phys.\ D {\bf 5} (1996) 813
  doi:10.1142/S0218271896000497
  [gr-qc/9612036].

\bibitem{Novello:2008ra}
  M.~Novello and S.~E.~P.~Bergliaffa,
  Phys.\ Rept.\  {\bf 463} (2008) 127
  doi:10.1016/j.physrep.2008.04.006
  [arXiv:0802.1634 [astro-ph]].

\bibitem{Battefeld:2014uga}
  D.~Battefeld and P.~Peter,
  Phys.\ Rept.\  {\bf 571} (2015) 1
  doi:10.1016/j.physrep.2014.12.004
  [arXiv:1406.2790 [astro-ph.CO]].

\bibitem{Buchbinder:1992rb}
  I.~L.~Buchbinder, S.~D.~Odintsov and I.~L.~Shapiro,
  Bristol, UK: IOP (1992) 413 p

\bibitem{Vilkovisky:1992pb}
  G.~A.~Vilkovisky,
  Class.\ Quant.\ Grav.\  {\bf 9} (1992) 895.
  doi:10.1088/0264-9381/9/4/008

\bibitem{Sotiriou:2008rp}
  T.~P.~Sotiriou and V.~Faraoni,
  Rev.\ Mod.\ Phys.\  {\bf 82} (2010) 451
  doi:10.1103/RevModPhys.82.451
  [arXiv:0805.1726 [gr-qc]].

\bibitem{DeFelice:2010aj}
  A.~De Felice and S.~Tsujikawa,
  Living Rev.\ Rel.\  {\bf 13} (2010) 3
  doi:10.12942/lrr-2010-3
  [arXiv:1002.4928 [gr-qc]].

\bibitem{Nojiri:2017ncd} 
  S.~Nojiri, S.~D.~Odintsov and V.~K.~Oikonomou,
  Phys.\ Rept.\  {\bf 692}, 1 (2017)
  doi:10.1016/j.physrep.2017.06.001
  [arXiv:1705.11098 [gr-qc]].
  
\bibitem{Nojiri:2010wj} 
  S.~Nojiri and S.~D.~Odintsov,
  Phys.\ Rept.\  {\bf 505}, 59 (2011)
  doi:10.1016/j.physrep.2011.04.001
  [arXiv:1011.0544 [gr-qc]].

  
\bibitem{ruz2}
 T. Ruzmaikina and A. Ruzmaikin,
 Sov. Phys. JETP {\bf 30}, 372 (1970).

\bibitem{Cai:2011bs} 
  Y.~F.~Cai and E.~N.~Saridakis,
  J.\ Cosmol.\  {\bf 17}, 7238 (2011)
  [arXiv:1108.6052 [gr-qc]].

\bibitem{Saridakis:2018fth} 
  E.~N.~Saridakis, S.~Banerjee and R.~Myrzakulov,
  Phys.\ Rev.\ D {\bf 98}, no. 6, 063513 (2018)
  doi:10.1103/PhysRevD.98.063513
  [arXiv:1807.00346 [gr-qc]].

\bibitem{Paul:2014cxa}
  N.~Paul, S.~N.~Chakrabarty and K.~Bhattacharya,
  JCAP {\bf 1410} (2014) no.10,  009
  doi:10.1088/1475-7516/2014/10/009
  [arXiv:1405.0139 [gr-qc]].

\bibitem{Bhattacharya:2015nda}
  K.~Bhattacharya and S.~Chakrabarty,
  JCAP {\bf 1602} (2016) no.02,  030
  doi:10.1088/1475-7516/2016/02/030
  [arXiv:1509.01835 [gr-qc]].

\bibitem{Bamba:2013fha}
  K.~Bamba, A.~N.~Makarenko, A.~N.~Myagky, S.~Nojiri and S.~D.~Odintsov,
  JCAP {\bf 1401} (2014) 008
  doi:10.1088/1475-7516/2014/01/008
  [arXiv:1309.3748 [hep-th]].

\bibitem{Carloni:2005ii}
  S.~Carloni, P.~K.~S.~Dunsby and D.~M.~Solomons,
  Class.\ Quant.\ Grav.\  {\bf 23} (2006) 1913
  doi:10.1088/0264-9381/23/6/006
  [gr-qc/0510130].

\bibitem{Elizalde:2010ts} 
  E.~Elizalde, S.~Nojiri, S.~D.~Odintsov, L.~Sebastiani and S.~Zerbini,
  Phys.\ Rev.\ D {\bf 83}, 086006 (2011)
  doi:10.1103/PhysRevD.83.086006
  [arXiv:1012.2280 [hep-th]].

\bibitem{Oikonomou:2018npe} 
  V.~K.~Oikonomou,
  Phys.\ Rev.\ D {\bf 97}, no. 6, 064001 (2018)
  doi:10.1103/PhysRevD.97.064001
  [arXiv:1801.03426 [gr-qc]].

\bibitem{Oikonomou:2013rba} 
  V.~K.~Oikonomou,
  Gen.\ Rel.\ Grav.\  {\bf 45}, 2467 (2013)
  doi:10.1007/s10714-013-1597-7
  [arXiv:1304.4089 [gr-qc]].

\bibitem{Cai:2016hea} 
  Y.~F.~Cai, A.~Marciano, D.~G.~Wang and E.~Wilson-Ewing,
  Universe {\bf 3}, no. 1, 1 (2016)
  doi:10.3390/universe3010001
  [arXiv:1610.00938 [astro-ph.CO]].

\bibitem{Cai:2014bea} 
  Y.~F.~Cai,
  Sci.\ China Phys.\ Mech.\ Astron.\  {\bf 57}, 1414 (2014)
  doi:10.1007/s11433-014-5512-3
  [arXiv:1405.1369 [hep-th]].

  
\bibitem{Barrow:1988xh}
  J.~D.~Barrow and S.~Cotsakis,
  Phys.\ Lett.\ B {\bf 214} (1988) 515.
  doi:10.1016/0370-2693(88)90110-4
  
\bibitem{Matsumoto:2013sba} 
  J.~Matsumoto,
  Phys.\ Rev.\ D {\bf 87}, no. 10, 104002 (2013)
  doi:10.1103/PhysRevD.87.104002
  [arXiv:1303.6828 [hep-th]].
  
\bibitem{Bertacca:2011wu} 
  D.~Bertacca, N.~Bartolo and S.~Matarrese,
\href{http://iopscience.iop.org/1475-7516/2012/08/021}{JCAP {\bf 1208}, 021 (2012)}
\href{http://arxiv.org/abs/1109.2082}{[arXiv:1109.2082 [astro-ph.CO]]}.

\bibitem{Mukhanov:1990me} 
  V.~F.~Mukhanov, H.~A.~Feldman and R.~H.~Brandenberger,
\href{http://www.sciencedirect.com/science/article/pii/037015739290044Z}{Phys.\ Rept.\  {\bf 215}, 203 (1992)}.

\bibitem{Tsujikawa:2009ku} 
  S.~Tsujikawa, R.~Gannouji, B.~Moraes and D.~Polarski,
\href{http://journals.aps.org/prd/abstract/10.1103/PhysRevD.80.084044}{Phys.\ Rev.\ D {\bf 80}, 084044 (2009)}
\href{http://arxiv.org/abs/0908.2669}{[arXiv:0908.2669 [astro-ph.CO]]}.
  
\bibitem{Bardeen:1980kt} 
  J.~M.~Bardeen,
  Phys.\ Rev.\ D {\bf 22}, 1882 (1980).
  doi:10.1103/PhysRevD.22.1882
  
\bibitem{Martin:2003sf} 
  J.~Martin and P.~Peter,
\href{http://journals.aps.org/prd/abstract/10.1103/PhysRevD.68.103517}{Phys.\ Rev.\ D {\bf 68}, 103517 (2003)} [\href{http://arxiv.org/abs/hep-th/0307077}{arXiv:hep-th/0307077}].

\bibitem{Cai:2012va} 
  Y.~F.~Cai, D.~A.~Easson and R.~Brandenberger,
  JCAP {\bf 1208}, 020 (2012)
  doi:10.1088/1475-7516/2012/08/020
  [arXiv:1206.2382 [hep-th]].


\bibitem{Kobayashi:2016xpl} 
  T.~Kobayashi,
  Phys.\ Rev.\ D {\bf 94}, no. 4, 043511 (2016)
  doi:10.1103/PhysRevD.94.043511
  [arXiv:1606.05831 [hep-th]].

\bibitem{Cai:2016thi} 
  Y.~Cai, Y.~Wan, H.~G.~Li, T.~Qiu and Y.~S.~Piao,
  JHEP {\bf 1701}, 090 (2017)
  doi:10.1007/JHEP01(2017)090
  [arXiv:1610.03400 [gr-qc]].

\bibitem{Raveendran:2017vfx} 
  R.~N.~Raveendran, D.~Chowdhury and L.~Sriramkumar,
  JCAP {\bf 1801}, no. 01, 030 (2018)
  doi:10.1088/1475-7516/2018/01/030
  [arXiv:1703.10061 [gr-qc]].
  
\bibitem{Odintsov:2017qif}
  S.~D.~Odintsov, D.~Sáez-Chillón Gómez and G.~S.~Sharov,
  Eur.\ Phys.\ J.\ C {\bf 77} (2017) no.12,  862
  doi:10.1140/epjc/s10052-017-5419-z
  [arXiv:1709.06800 [gr-qc]].


\bibitem{Schmidt:1998sn}
  H.~J.~Schmidt,
  gr-qc/9808060.

\bibitem{Wei:2015xax}
  H.~Wei, H.~Y.~Li and X.~B.~Zou,
  Nucl.\ Phys.\ B {\bf 903} (2016) 132
  doi:10.1016/j.nuclphysb.2015.12.006
  [arXiv:1511.00376 [gr-qc]].

\bibitem{Bhattacharya:2017evz} 
  P.~Bari, P.~Bari, S.~Chakraborty and K.~Bhattacharya,
  Gen.\ Rel.\ Grav.\  {\bf 50}, no. 9, 118 (2018)
  doi:10.1007/s10714-018-2443-8
  [arXiv:1711.11395 [gr-qc]].

\bibitem{Amendola:2006we} 
  L.~Amendola, R.~Gannouji, D.~Polarski and S.~Tsujikawa,
  Phys.\ Rev.\ D {\bf 75}, 083504 (2007)
  doi:10.1103/PhysRevD.75.083504
  [gr-qc/0612180].

\bibitem{Carloni:2007br} 
  S.~Carloni, A.~Troisi and P.~K.~S.~Dunsby,
  Gen.\ Rel.\ Grav.\  {\bf 41}, 1757 (2009)
  doi:10.1007/s10714-008-0747-9
  [arXiv:0706.0452 [gr-qc]].

\bibitem{Carloni:2004kp} 
  S.~Carloni, P.~K.~S.~Dunsby, S.~Capozziello and A.~Troisi,
  Class.\ Quant.\ Grav.\  {\bf 22}, 4839 (2005)
  doi:10.1088/0264-9381/22/22/011
  [gr-qc/0410046].

\bibitem{Odintsov:2017tbc} 
  S.~D.~Odintsov and V.~K.~Oikonomou,
  Phys.\ Rev.\ D {\bf 96}, no. 10, 104049 (2017)
  doi:10.1103/PhysRevD.96.104049
  [arXiv:1711.02230 [gr-qc]].
\end{thebibliography}
\end{document}